%% file: main.tex
\documentclass[dvips]{llncs} 
\usepackage{wrapfig}
\usepackage{graphics}
\usepackage{subfigure}
\usepackage{color}
\usepackage{epsfig}

\usepackage[T1]{fontenc}
\usepackage{amsmath}
\usepackage{amssymb}
\usepackage{latexsym}
\usepackage{wasysym}
\usepackage{bussproofs}

\usepackage{array}
\usepackage{alltt}
\usepackage{theorem}

\title{CCS for Trees}
\author{Thomas Ehrhard\inst{1} \and Ying Jiang\inst{2}}
\institute{
CNRS, PPS, UMR 7126,\\
Univ Paris Diderot, 
Sorbonne Paris Cité, F-75205 Paris, France
\email{thomas.ehrhard@pps.univ-paris-diderot.fr}
\and
State Key Laboratory of Computer Science\\
Institute of Software, Chinese Academy of Sciences\\
P.O. Box 8718, 100190 Beijing, China\\
\email{jy@ios.ac.cn}
}

\input notation

\begin{document}
\maketitle

\begin{abstract}
  CCS can be considered as a most natural extension of finite state automata in
  which \emph{interaction} is made possible thanks to parallel composition.
  We propose here a similar extension for top-down tree automata. We introduce
  a parallel composition which is parameterized by a graph at the vertices of
  which subprocesses are located. Communication is allowed only between
  subprocesses related by an edge in this graph. We define an observational
  equivalence based on barbs as well as weak bisimilarity equivalence and prove
  an adequacy theorem relating these two notions.
\end{abstract}

\section*{Introduction}
There is no need to insist on the importance of tree automata~
\cite{ComonAndAl2007} in modern theoretical and applied computer science: they
are pervasive in logic, verification, rewriting, structured documents handling,
constraint solving etc. Tree automata are similar to usual finite word automata
with the difference that they recognize trees instead of words (sequences of
letters). Let $\Sigma$ be a ranked signature ($\Sigma_n$ is the set of function
symbols of arity $n$). A $\Sigma$-tree is just a term written with the
signature $\Sigma$. A \emph{top-down tree automaton} has a finite number of
\emph{states} and transitions labeled by elements of $\Sigma$: a transition
labeled by $f\in\Sigma_n$ has a \emph{source} and a sequence of $n$
\emph{targets} which all are states of the automaton. A word automaton can be
seen as a tree automaton over a signature $\Sigma$ such that $\Sigma_n$ is
empty for all $n>1$ and $\Sigma_0$ has a unique distinguished element $*$.

The definition of tree recognition by a top-down tree automaton $A$ is quite
simple: a tree $f(t_1,\dots,t_n)$ is recognized by $A$ at state $X$ means that
$A$ has an $f$-labeled transition whose source is $X$ and target is $(\List
X1n)$ and $t_i$ is recognized by $A$ at state $X_i$ for each
$i=1,\dots,n$. There is also a notion of bottom-up tree automata, that we do
not consider in this work; these two notions are equivalent in terms of the
recognized languages, as long as one considers \emph{non-deterministic}
automata.

Automata feature a \emph{dualist} vision of computation with an essential
dichotomy between programs (automata) and data (words, trees), very much in the
spirit of Turing machines (based on the machine/tape dichotomy). The process
algebra \CCS{}, introduced in the early 1980's by Milner~\cite{Milner80},
encompasses this restriction, extending finite automata with interactive
capabilities. In this framework, finite automata (labeled with letters
$a,b,\dots$) can typically interact with other automata (labeled with dual
letters $\overline a,\overline b,\dots$), as soon as they are combined through
a new binary operation: \emph{parallel composition}. But much more general
interaction scenarii are of course possible in \CCS{}. This fundamental
invention led to very fruitful new lines of research in the theory of
concurrent processes and to the introduction of new process algebra, among
which the $\pi$-calculus~\cite{MilnerParrowWalker92a} is not the less
remarkable, with many spectacular applications to cryptography, bioinformatics
etc.

In this paper, we propose a similar ``interactive closure'' of tree automata, a
new version of \CCS{} which extends tree automata just as ordinary \CCS{}
extends word automata. 

The natural idea is of course to add a parallel composition operation on
processes, but this requires some care. Indeed when a prefixed process $\PREF
f{(\List P1n)}$ ---~after a prefix $f\in\Sigma_n$, it is natural to have $n$
subprocesses, and not only one, as explained in~\cite{ChaiQuJiang08}~---
interacts with a dually prefixed one $\PREFO f{(\List Q1n)}$, we should remove
the prefixes (just as in \CCS{}) and then authorize interaction between the
subprocess $P_i$ with all processes which could communicate with its father
$\PREF f{(\List P1n)}$ as well as with $Q_i$, \emph{but not with the $Q_j$'s
  for $j\not=i$}; neither should the $P_i$'s be allowed to communicate with
each other in the resulting process. The same should hold of course for the
$Q_i$'s.

One major motivation for this choice of design is that top-down tree
recognition of tree automata should be implementable in our new \CCS{} for
trees, just as usual word recognition of automata is implementable in ordinary
\CCS{}. But for this purpose we have to preserve carefully the distinction
between the various sons of tree nodes, thus preventing sons which are not at
similar positions to interact. Indeed, with this definition, we are able to
prove the interactive recognition Theorem~\ref{th:tree-recognition}.

This led us to the idea that general parallel composition should be a
\emph{graph}, at the vertices of which subprocesses (which are guarded sums)
should be located; the edges of this graph specify which interactions are
allowed. In Section~\ref{sec:syntax}, we introduce the syntax of this new
process calculus \TCCS, restricting ourselves to a fragment where all sums are
guarded; indeed, the corresponding fragment of \CCS{} is known to be sensible
and well behaved.

In Section~\ref{sec:operational}, we introduce an operational semantics for
\TCCS{} by defining a single rewriting rule. This rule generalizes the
$a/\overline a$ reduction of \CCS{} to the case where $a$ can be an $n$-ary
function symbol and implements the idea of restricted communication
capabilities explained above. 

In order to define an operational equivalence on processes, we adapt the
concept of \emph{weak barbed
  congruence}~\cite{MilnerSangiorgi92,SangiorgiWalker01} which is a natural way
of saying that two processes behave in the same way, in all possible
contexts. As usual, this notion is quite difficult to handle and we introduce
therefore a notion of weak bisimilarity in Section~\ref{sec:wbisim} and prove
that two weakly bisimilar processes are weakly barbed congruent in
Section~\ref{sec:adequacy}. For this, we define a labeled transition system on
processes, and the definition of its transitions involves crucially the
locations (graph vertices). The notion of bisimulation itself has to take these
locations carefully into account.

In Section~\ref{sec:operational}, we also argue that our version of \CCS{} is a
conservative extension of both tree automata and ordinary \CCS{}: by this we
mean that it admits restrictions which coincide with these two
formalisms. Moreover, we show that tree recognition can be expressed simply in
terms of interaction, using only the rewriting semantics. Though quite simple,
this result uses in an essential way the restricted communication capabilities
of \TCCS.

These results suggest that \TCCS{} is a sound and interesting extension of
\CCS{}. The most novel feature is that subprocesses are located at the vertices
of a graph whose edges indicate which communications are possible, and the
topology of this graph evolves during reduction. When no edge relates two
processes, they can evolve independently, in a truly concurrent way, whereas
the presence of an edge means that the corresponding processes will possibly
synchronize in the future. Another interesting property of this approach is the
importance of \emph{locations} which suggests connections with the work of
Castellani~\cite{Castellani01}, though locations are used in a different way:
in this latter work, communication is possible when the involved processes are
located at the same place.

This paper extends non trivially~\cite{ChaiQuJiang08}, where parallel
composition however was not dealt with. Finding the right way of formalizing
this operation and of defining the relevant notions of bisimulation have been a
difficult task. Beyond the interactive closure of tree automata obtained by
this new formalism, we also believe that \TCCS{} provides a new compositional
framework for the study of true concurrency. Indeed, the $n$ processes forked by
an $n$-ary labeled prefix behave in a truly concurrent way, and such a truly
concurrent situation cannot be obtained in ordinary \CCS{} (concurrency is
modelized by interleaving).

One of our further works will deal with possible connections between \TCCS{}
and other process algebras, and in particular with the possibility of encoding
\TCCS{} within the $\pi$-calculus.

\section{Syntax of processes}\label{sec:syntax}
We use letters $\vec P,\vec Q,\dots$ to denote vectors $(\List P1n)$, $(\List
Q1n)$ etc. Let $\Locations$ be a countable set whose elements are called
\emph{locations} denoted with letters $p,q\dots$ with or without subscripts or
superscripts.

\subsection{Graphs}
Let $E$ and $F$ be disjoint sets and let $p\in E$. We set $\Subst
EFp=(E\setminus\{p\})\cup F$. In other words, $\Subst EFp$ is the set obtained
from $E$ by substituting the element $p$ with the set $F$. 

By a graph we mean a pair $G=(\Web G,\Coh G)$, where $\Web G$ is a finite
subset of $\Locations$ and $\Coh G$ is a symmetric and antireflexive relation
on $\Web G$. Let $G$ and $H$ be graphs with $\Web G\cap\Web H=\emptyset$ and
let $p\in\Web G$. We define a graph $\Subst GHp$ as follows:
\begin{itemize}
\item $\Web{\Subst GHp}=\Subst{\Web G}{\Web H}{p}$
\item and, given $q,r\in\Web{\Subst GHp}$, we say that $q\Rel{\Coh{\Subst
      GHp}}r$ if $q\Rel{\Coh G}r$ or $q\Rel{\Coh H}r$ or $q\Rel{\Coh G}p$ and
  $r\in\Web H$ or $r\Rel{\Coh G}p$ and
  $q\in\Web H$.
\end{itemize}

\subsection{Processes}

We assume to be given a countable set of \emph{processes variables}
$\Variables$, denoted with letters $X,Y,\dots$ with or without subscripts or
superscripts.

Let $\Sigma=(\Sigma_n)_{n\in\Nat}$ be a signature. With any symbol
$f\in\Sigma_n$, we associate a \emph{co-symbol} $\Cosymb f$ distinct from all
the elements of $\Sigma_n$ and we set $\Barset{\Sigma}_n=\Sigma_n\cup\{\Cosymb
f\St f\in\Sigma_n\}$.  In that way, we define an extended signature
$\Barset\Sigma=(\Barset\Sigma_n)_{n\in\Nat}$. For $f\in\Sigma_n$, we set
$\Cosymb{\Cosymb f}=f$.

We define the set of $\TCCS$ processes by induction.
\begin{itemize}
\item If $X\in\Variables$ then $X$ is a process.
\item If $X\in\Variables$ and $P$ is a process, then $\FIX XP$ is a process in
  which $X$ is bound.
\item If $f\in\Barset\Sigma_n$ and $\List P1n$ are processes, then $\PREF
  f{(\List P1n)}$ is a process.
\item If $G$ is a finite $\Locations$-graph (that is $\Web
  G\subseteq\Locations$ is finite) and $\Phi$ is a function from $\Web G$ to
  processes, then $\PAR G\Phi$ is a process, to be understood as the parallel
  composition of the processes $\Phi(p)$ for $p\in\Web G$, with communication
  capabilities specified by $G$. The processes $\Phi(p)$ are called the
  \emph{components} of $\PAR G\Phi$.
\item $\ZERO$ is a process and if $P$ and $Q$ are processes, then $P\PLUS Q$ is
  a process.
\item If $P$ is a process and $I$ is a finite subset of $\Sigma$, then $\RESTR
  PI$ is a process.
\end{itemize}

The notion of free and bound variable does not deserve further comments, $\mu$
being of course a binder.

\subsection{$\alpha$-conversions of locations.} 
Two processes $P$ and $P'$ such that there exists a bijection $\phi:\Web
P\to\Web{P'}$ which is a graph isomorphism (that is $p\Coh
Pq\Equiv\phi(p)\Coh{P'}\phi(q)$) and $P'(\phi(p))=P(p)$ for all $p\in\Web P$
are said to be \emph{externally $\alpha$-equivalent}. General
$\alpha$-equivalence is defined by extending this relation to sub-processses in
the obvious way.

When we consider several processes $\List P1n$ at the same time, we always
assume that the webs $\Web{P_1},\dots,\Web{P_n}$ are pairwise disjoint.

\subsection{Substitution.}
If $R$ and $P$ are processes and $X\in\Variables$, then the process $\Subst
RPX$ is defined in the obvious way, substituting each occurrence of $X$ in $R$
with $P$. Of course, one has as usual to perform $\alpha$-conversion when needed
during this process.


\subsection{Canonical processes}
We define now the notion of \emph{canonical process}: it is a process where all
sums are guarded. More precisely, we define by mutual induction three classes
of objects:
\begin{itemize}
\item canonical processes,
\item \emph{canonical guarded sum}
\item and \emph{recursive canonical guarded sum}.
\end{itemize}
These are particular processes on which we'll focuss our attention in the
sequel. 

\begin{itemize}
\item If $X\in\Variables$ then $X$ is a canonical process.
\item If $G$ is a finite $\Locations$-graph and $\Phi$ is a function from $\Web
  G$ to recursive canonical guarded sums, then $\PAR G\Phi$ is a canonical
  process.
\item If $P$ is a canonical process and $I$ is a finite subset of $\Sigma$,
  then $\RESTR PI$ is a canonical process.
\item A canonical guarded sum is either $\ZERO$ or a process of the shape
  $\PREF f{(\List P1n)}\PLUS S$ where $f\in\Barset\Sigma_n$, $S$ is a canonical
  guarded sum and $\List P1n$ are canonical processes.
\item A recursive canonical guarded sum is either a canonical guarded
  sum or a process of shape $\FIX XS$ where $S$ is a recursive canonical 
  guarded sum.
\end{itemize}

For instance, the processes $\PAR G\Phi+\PAR H\Psi$ and $\FIX XX$ are not
canonical.

\begin{lemma}\label{lemma:canonical-substitution}
  Let $R$ and $P$ be canonical processes. Then $\Subst RPX$ is a canonical
  process. If $R$ is a recursive canonical guarded sum, then so is $\Subst
  RPX$. If $R$ is a canonical guarded sum, then so is $\Subst RPX$.
\end{lemma}
\Beginproof
Easy induction on $R$.
\Endproof

With any \emph{recursive canonical guarded sum} $S$, we associate a
\emph{canonical guarded sum} $\Cansum S$ as follows:
\begin{equation*}
\Cansum S=
\begin{cases}
  S & \text{if $S$ is a canonical guarded sum}\\
  \Cansum{\Subst TSX} & \text{if $S=\FIX XT$}\,.
\end{cases}
\end{equation*}

Using Lemma~\ref{lemma:canonical-substitution}, one sees easily that this
function is well defined and total.

\emph{All the processes we consider in this paper are canonical}. By
Lemma~\ref{lemma:canonical-substitution}, processes are closed by substitution.

We denote with $\Processes$ the set of all canonical processes.  If $P=\PAR
G\Phi$ is a canonical process, we use $\Locof P=\Web G$. Also, for $p\in\Web
P$, we often write $P(p)$ instead of $\Phi(p)$, and we denote as $\Coh P$ the
graph relation of $G$.

The empty process (the only $P$ such that $\Web P=\emptyset$) is denoted as
$\EMPTY$.

\subsection{More notations}\label{sec:notations}
Given two graphs $G$ and $H$ with disjoint webs, and a subset $D$ of $\Web
G\times\Web H$ we define a graph $K=G\Gplus DH$ by $\Web{K}=\Web G\cup\Web H$
and, given $p,q\in\Web K$, we stipulate that $p\Coh Kq$ if $p\Coh Gq$ or $p\Coh
Hq$ or $(p,q)\in D$ or $(q,p)\in D$. If $D=\emptyset$ then we set $\Plus
GH=G\Gplus DH$.

Given processes $P=\PAR G\Phi$ and $Q=\PAR H\Psi$ and a relation
$D\subseteq\Web P\times\Web Q$, one defines the process $P\Gplus DQ$ as
$\PAR{(G\Gplus DH)}{\Phi\cup\Psi}$. When $D$ is empty we simply denote this sum
as $P\IPlus Q$, and more generally, we denote as $\IPlus\vec P$ the sum
$P_1\IPlus\cdots\IPlus P_n$ of the processes $\vec P=(\List P1n)$ (remember that
we implicitly assume that the sets $\Web{P_i}$ are pairwise disjoint). When
$D=\Web P\times\Web Q$, the process $P\Gplus DQ$ will be denoted as $\PARFULL
PQ$ and called the \emph{full parallel composition} of $P$ and $Q$. It
corresponds to the standard parallel composition of process algebras, where all
processes can freely interact with each other.

With the same notations as above, if $p\in\Web G$, we denote as $\Subst PQp$
the process $\PAR{\Subst GHp}{\Phi'}$ where $\Phi'(p')=\Phi(p')$ if
$p'\notin\Web H$ and $\Phi'(p')=\Psi(p')$ if $p'\in\Web H$. 

\section{Operational semantics}\label{sec:operational}

\subsection{Internal reduction}\label{sec:internal-reduction}

Let $P$ and $P'$ be processes. We say that $P$ \emph{reduces} to $P'$ if there
are $p,q\in\Web P$ such that $p\Coh Pq$, $\Cansum{P(p)}=\PREFI f{(\List
  P1n)}+S$, $\Cansum{P(q)}=\PREFO f{(\List Q{1}{n})}+T$ and $P'$ is defined as
follows\footnote{We heavily use the implicit hypothesis that, when several
  processes $\List P1n$ are considered at the same time, the sets $\Web{P_i}$
  are pairwise disjoint.}: $\Web{P'}=(\Web
P\setminus\{p,q\})\cup\bigcup_{i=1}^n\Web{P_i}\cup\bigcup_{i=1}^n\Web{Q_i}$ and
$\Coh{P'}$ is the least symmetric relation on $\Web{P'}$ such that, for any,
$p',q'\in\Web{P'}$, one has $p'\Coh{P'}q'$ in one of the following cases:
\begin{enumerate}
\item $p'\Coh{P_i}q'$ or $p'\Coh{Q_i}q'$ for some $i=1,\dots,n$
\item $p'\in\Web{P_i}$ and $q'\in\Web{Q_i}$ for some $i=1,\dots,n$
  (\emph{the same $i$ for both})
\item
  $\{p',q'\}\not\subseteq\bigcup_{i=1}^n\Web{P_i}\cup\bigcup_{i=1}^n\Web{Q_i}$
  and $\lambda_1(p')\Coh P\lambda_1(q')$
\end{enumerate}
where $\lambda_1:\Web{P'}\to\Web P$ is the \emph{residual function} defined by
\begin{equation*}
  \lambda_1(p')=
  \begin{cases}
    p & \text{if $p'\in\bigcup_{i=1}^n\Web{P_i}$}\\    
    q & \text{if $p'\in\bigcup_{i=1}^n\Web{Q_i}$}\\
    p' & \text{otherwise.}
  \end{cases}
\end{equation*}
Observe that $\lambda_1$ is not a surjection when $n=0$.

We finish the definition of $P'$ by saying that $P'(p')=P_i(p')$ if
$p'\in\Web{P_i}$, $P'(p')=Q_i(p')$ if $p'\in\Web{Q_i}$ (for $i=1,\dots,n$) and
$P'(p')=P(p')$ if
$p'\notin\bigcup_{i=1}^n\Web{P_i}\cup\bigcup_{i=1}^n\Web{Q_i}$.

This crucial definition clearly deserves some explainations. The process $P$ to
be reduced has two subprocesses located at $p$ and $q$, with dual prefixes:
$\PREF f{\vec P}$ and $\PREFO f{\vec Q}$. The fact that $p$ and $q$ are
connected in $P$ ($p\Coh Pq$) means that these processes can interact. This
interaction consists in suppressing both prefixes and in replacing the vertice
$p$ of the graph $G$ of $P$ by the graph $G_1\IPlus\cdots\IPlus G_n$ (where
$G_i$ is the graph of $P_i$) and the vertice $q$ by the graph
$H_1\IPlus\cdots\IPlus H_n$ (where $H_i$ is the graph of $Q_i$) within the
graph $G$ of $P$. The connection between $p$ and $q$ in $P$ is inherited by the
vertices of $G_i$ and $H_i$ in $P'$, but a process located on $G_i$ (one of the
components of $P_i$) cannot communicate with a process located on $H_j$ with
$j\not=i$. The connections between $p$ and other vertices of $P$, distinct from
$q$, are also inherited by the vertices of all $G_i$'s and similarly for the
$H_i$'s.

We denote with $\Intred$ the internal reduction relation and with $\Intredtr$
its reflexive and transitive closure.

\begin{example}
  Let $a\in\Sigma_0$ and $f\in\Sigma_2$. Consider the process $P=\Coname a\mid
  a\mid\PREF f{(a,\Coname a)}\mid\PREFO f{(a,\Coname a)}$ (we write simply
  ``$a$'' instead of $\PREF a{()}$). In other words, the graph of $P$ is a
  complete graph with $4$ vertices, say $1,2,3,4$, and we have $P(1)=a$,
  $P(2)=\Coname a$, $P(3)=\PREF f{(a,\Coname a)}$ and $P(4)=\PREFO f{(a,\Coname
    a)}$. Since $3$ and $4$ are connected in that graph and the corresponding
  prefixes $f$ and $\Coname f$ are dual, we can reduce $P$ to a process $P'$
  such that $\Web{P'}=\{1,2,5,6,7,8\}$ (remember that we work up to
  $\alpha$-equivalence, so the names of locations are irrelevant) with
  $P'(1)=a$, $P'(2)=\Coname a$, $P'(5)=a$, $P'(6)=\Coname a$, $P'(7)=a$, and
  $P'(8)=\Coname a$, and the edges of $P'$ are all $\{i,j\}$ with $i\in\{1,2\}$
  and $j\not=i$, $\{5,7\}$ and $\{6,8\}$. So, in $P'$, the interaction of $a$
  located at $5$ with $\Coname a$ located at $8$ is not possible, but of
  course $a$ located at $5$ can interact with $\Coname a$ located at
  $2$. Performing that reduction, we get $P''$ with $\Web{P''}=\{1,6,7,8\}$ and
  the edges of $P''$ are all $\{1,j\}$ with $j\not=1$ and $\{6,8\}$, with
  $P''(1)=a$, $P''(6)=\Coname a$, $P''(7)=a$ and $P''(8)=\Coname a$. In $P''$,
  the only possible reductions are between $a$ located at $1$ and $\Coname a$
  located at $6$ or $8$. Both lead to the process $a\IPlus\Coname a$ where no
  reduction is possible.
\end{example}

\subsection{Top-down tree automata as a particular case}
\label{sec:automata-as-processes}

A \emph{top-down tree automaton} is a pair $A=(\cQ,\cT)$ where $\cQ$ is a
finite subset of $\Variables$, whose elements are called \emph{states}, and
$\cT$ is a finite set of triples $(X,f,(\List X1n))$ where $f\in\Sigma_n$ and
$\List X1n\in\cQ$ and whose elements are called \emph{transitions}. The
\emph{language recognized by $A$ at state $X\in\cQ$}, denoted as $\Lang AX$, is
the least set of $\Sigma$-trees such that $f(\List t1n)\in\Lang AX$ as soon as
there are $\List X1n\in\cQ$ such that $(X,f,(\List X1n))\in\cT$ and
$t_i\in\Lang{A}{X_i}$ for $i=1,\dots,n$.

We associate a process $\ProcofTA AX$ with any pair $(A,X)$ where $A=(\cQ,\cT)$
is a tree automaton and $X\in\cQ$. More generally we define $\ProcofTA AX^\cX$
where $\cX$ is a finite subset of $\Variables$ (intuitively, $\cX$ is the set
of already defined processes), and then we set $\ProcofTA AX=\ProcofTA
AX^\emptyset$.
\begin{itemize}
\item If $X\notin\cX$, then $\ProcofTA AX^\cX=\FIX XS$ where $S$ is the sum of
  all prefixed processes $\PREF f{(\ProcofTA
    A{X_1}^{\cX\cup\{X\}},\dots,\ProcofTA A{X_n}^{\cX\cup\{X\}})}$ where
  $(X,f,(\List X1n))\in\cT$,
\item and if $X\in\cX$, then $\ProcofTA AX^\cX=X$.
\end{itemize}
This inductive definition is well founded because the parameter $\cX$ increases
strictly at each inductive step, and remains included in $\cQ$. Moreover, the
invariant that all the free variables of $\ProcofTA AX^\cX$ belong to
$\cX$ is preserved by the inductive step, and hence $\ProcofTA AX$ is closed.

\begin{lemma}\label{lemma:fix-TA-exp}
  With the notations above, $\Cansum{\ProcofTA AY}$ is the sum of all prefixed
  processes $\PREF f{(\ProcofTA A{Y_1},\dots,\ProcofTA A{Y_n})}$ where
  $(Y,f,(\List Y1n))\in\cT$.
\end{lemma}
\Beginproof
More generally,
$\Cansum{\Substbis{\ProcofTA AX^{\{X_1,\dots,X_p\}}}{\ProcofTA
    A{X_1}/X_1,\dots,\ProcofTA A{X_p}/X_p}}$ is equal to the sum above, for any
subset $\{X_1,\dots,X_p\}$ of $\cQ$ (with the $X_i$'s pairwise distinct). The
proof is a simple induction on $q-p$, where $q$ is the cardinality of $\cQ$.
\Endproof

We represent dually any $\Sigma$-tree $t=f(t_1,\dots,t_n)$ as a process
$\ProcofT t$ by setting $\ProcofT t=\PREFO
f{(\ProcofT{t_1},\dots,\ProcofT{t_n})}$. The following results expresses that
our process algebra, together with its internal reduction, is a conservative
extension of tree automata by showing that tree recognition boils down to a
(very) particular case of interaction between processes.

\begin{theorem}\label{th:tree-recognition}
  Let $A=(\cQ,\cT)$ be a tree automaton, let $X\in\cQ$ and let $t$ be a
  $\Sigma$-tree. Then $t\in\Lang AX$ iff $(\PARFULL{\ProcofTA AX}{\ProcofT
    t})\Rel\Intredtr\EMPTY$.
\end{theorem}
\Beginproof
This is straightforward, once observed that, if $t=f(\List t1n)$ and if
$(X,f,(\List X1n))\in\cT$, one has $\PARFULL{\ProcofTA AX}{\ProcofT t}\Rel
\Intred(\PARFULL{\ProcofTA
  A{X_1}}{\ProcofT{t_1}})\IPlus\cdots\IPlus(\PARFULL{\ProcofTA
  A{X_n}}{\ProcofT{t_n}})$, thanks to Lemma~\ref{lemma:fix-TA-exp}. Observe
then that $(\PARFULL{\ProcofTA
  A{X_1}}{\ProcofT{t_1}})\IPlus\cdots\IPlus(\PARFULL{\ProcofTA
  A{X_n}}{\ProcofT{t_n}})$ reduces to $\EMPTY$ iff each process
$\PARFULL{\ProcofTA A{X_i}}{\ProcofT{t_i}}$ reduces to $\EMPTY$ since these
processes cannot interact with each other. If $\cT$ has no element of the
shape $(X,f,(\List X1n))$, then the process $\PARFULL{\ProcofTA AX}{\ProcofT
    t}$ does not reduce. 
\Endproof

\subsection{\CCS{} for words as a particular case}
\label{sec:CCS-subsystem}
We assume here that $\Sigma_n=\emptyset$ for all $n>1$ and that
$\Sigma_0=\{*\}$. Then a $\Sigma$-tree is the same thing as a $\Sigma_1$-word,
written $a_1\dots a_p*$. We restrict our attention to processes in which all
the graphs parameterizing parallel compositions are complete, so that any
process is of the shape $S_1\mid\cdots\mid S_p$ where each $S_i$ is a recursive
canonical guarded sum $\FIX{\vec X}{(\PREF{a_1}{P_1}+\cdots+\PREF{
    a_m}{P_m})}$: this restriction of our process algebra coincides with
\emph{guarded \CCS{}}. Observe also that, if $P$ is a process in this
restricted setting (arities $\leq 1$ and all parallel compositions are complete
graphs), and if $P\Rel{\Intred}P'$, then $P'$ belongs to the same restriction
and the reduction $P\Rel{\Intred}P'$ is a standard $\tau$-reduction of
\CCS{}. In that way we see that our process algebra is also a conservative
extension of ordinary guarded \CCS{}.

There is a slight, innocuous, variation in this way of representing ordinary
\CCS{} within \TCCS. It consists in taking $\Sigma_n=\emptyset$ for $n\neq 1$
and $\Sigma_1$ as word alphabet. Then one can use $\EMPTY$ (the empty process)
instead of the $*$ symbol of arity $0$. For simplicity, it is this coding that
we'll use in Section~\ref{sec:CCS-bisimilarity}. The drawback of this
representation is that it does not scale down to automata considered as
particular processes as explained in Section~\ref{sec:automata-as-processes}.

\subsection{Weak barbed bisimilarity}

Let $f\in\Barset\Sigma$ and let $P$ be a process. We say that $f$ is a
\emph{barb} of $P$, and write $\Barb Pf$, if there exists $p\in\Web P$ such
that $\Cansum{P(p)}$ is of shape $\PREF f{(\List P1n)}+S$.

A relation $\cB\subseteq\Processes^2$ is a \emph{weak barbed bisimulation} if
it is symmetric and satisfies the following conditions. For any
$P,Q\in\Processes$ such that $P\Rel\cB Q$,
\begin{itemize}
\item for any $P'\in\Processes$, if $P\Rel\Intredtr P'$, then there exists
  $Q'\in\Processes$ such that $Q\Rel\Intredtr Q'$ and $P'\Rel\cB Q'$ (one says
  that $\cB$ is a \emph{weak reduction bisimulation});
\item for any $P'\in\Processes$ and any $f\in\Barset\Sigma$, if $P\Rel\Intredtr
  P'$ and $\Barb{P'}f$, then there exists $Q'\in\Processes$ such that
  $Q\Rel\Intredtr Q'$ and $\Barb{Q'}f$ (one says that $\cB$ is \emph{weak barb
    preserving}; observe that one does not require that $P'\Rel\cB Q'$).
\end{itemize}


The diagonal relation $\{(P,P)\St P\in\Processes\}$ is a weak barbed
bisimulation, and if $\cB$ and $\cB'$ are weak barbed bisimulations, then so
are $\cB'\Comp\cB$ and $\cB\cup\cB'$. We say that $P,Q\in\Processes$ are
\emph{weakly barbed bisimilar} if there exists a weak barbed bisimulation $\cB$
such that $P\Rel\cB Q$. Notation: $P\Rel{\Wbarbbis}Q$.

\begin{lemma}\label{lemma:Wbarbbis-equiv}
  Weak barbed bisimilarity is an equivalence relation.
\end{lemma}
\Beginproof
Straightforward, using the above closure properties of weak barbed
bisimulations.
\Endproof

\subsection{Weak barbed congruence}

Let $Y$ be a variable; a \emph{$Y$-context} is a process $R$ which contains
exactly one free occurrence of $Y$, which does not occur in a subprocess of $R$
of the shape $\FIX{X}{R'}$ (in other words, $Y$ must really occur only
once in $R$). If $R$ and $S$ are $Y$-contexts, so is $\Subst RSY$.

A relation $\cR\subseteq\Processes^2$ is a \emph{congruence} if it is reflexive
and such that, for any $Y$-context $R$, one has $P\Rel\cR Q\Implies \Subst
RPY\Rel\cR\Subst RQY$.

\begin{proposition}
  For any reflexive relation $\cR\subseteq\Processes^2$, there exists a largest
  congruence $\overline\cR$ contained in $\cR$. This relation is characterized
  by: $P\Rel{\overline\cR}Q$ iff for any $Y$-context $R$ one has $\Subst
  RPY\Rel\cR\Subst RQY$. If $\cR$ is an equivalence relation, so is
  $\overline\cR$.
\end{proposition}
\Beginproof
The first statement results from the fact that congruences are closed under
arbitrary unions and that $\cR$ contains the identity relation which is a
congruence. As to the second statement, let $\cE$ be the relation defined by
$P\Rel{\cE}Q$ iff for any $Y$-context $R$ one has $\Subst RPY\Rel\cR\Subst
RQY$. Then $\cE$ is a congruence which is contained in $\cR$ (since we can take
$R=Y$) and hence $\cE\subseteq\overline\cR$. Conversely, assume that
$P\Rel{\overline\cR}Q$ and let $R$ be a $Y$-context. Since $\overline\cR$ is a
congruence, we have $\Subst RPY\Rel{\overline\cR}\Subst RQY$ and hence $\Subst
RPY\Rel{\cR}\Subst RQY$ since $\overline\cR\subseteq\cR$ by definition of
$\overline\cR$ and hence $P\Rel\cE Q$. The last statement results from the
second one since $\cE$ is an equivalence relation when $\cR$ is an equivalence
relation.
\Endproof
The largest congruence contained in $\Wbarbbis$ is denoted as $\Wbarbcong$ and
is called \emph{weak barbed congruence}: it is our main notion of operational
equivalence on processes. It is an equivalence relation by the proposition
above and by Lemma~\ref{lemma:Wbarbbis-equiv}. Moreover, we have
\begin{equation*}
P\Rel\Wbarbcong Q \text{ iff for any $Y$-context $R$, we have } \Subst
RPY\Rel\Wbarbbis\Subst RQY\,.  
\end{equation*}

\section{Localized transition systems of processes}
\label{sec:wbisim}

Just as in ordinary \CCS{}, it is very difficult to prove that two processes
are weak barbed congruent, because of the universal quantification on contexts
used in the definition of this equivalence relation. In order to prove weak
barbed congruence of processes, one needs therefore more convenient tools.

The most canonical of these tools is \emph{weak bisimilarity}, an equivalence
relation which expresses that two processes manifest the same communication
capabilities along their internal reductions. This equivalence relation is
defined as the union of all \emph{weak bisimulations}.

The main feature of weak bisimilarity is that it is a congruence: this fact is
the main ingredient in the proof that two weakly bisimilar processes are weakly
barbed congruent. To prove this result, one needs to associate with each weak
bisimulation $\cR$ a new weak bisimulation $\cR'$ called its \emph{parallel
  extension}. In ordinary \CCS{}, the definition is as follows: one says that
$U\Rel{\cR'}V$ if $U=\PARFULL PS$ and $V=\PARFULL QS$ with $P\Rel\cR Q$ and $S$
is a process. The main step is of course to show that $\cR'$ is a weak
bisimulation.

In \TCCS{} however, we cannot simply speak of ``the parallel composition'' $U$
of $P$ and $S$, we have to specify a relation $C\subseteq\Web P\times\Web S$,
and then we can set $U=P\Gplus CS$. Similarly we have to say that $V=Q\Gplus
DS$ for some relation $D\subseteq\Web Q\times\Web S$, and that $P\Rel\cR
Q$. Not surprisingly, we shall see that these relations $C$ and $D$ must
fulfill some requirement.

Moreover our bisimulations cannot be simple relations between processes,
because, when two processes $P=\PAR G\Phi$ and $Q=\PAR H\Psi$ are bisimilar, we
have to say which subprocesse $\Phi(p)$ of $P$ should be in bisimulation with
which subprocesses $\Psi(q)$ of $Q$. 

For instance, if $P=\PREF f{(P_1,P_2)}$ and $Q=\PREF f{(Q_1,Q_2)}$ (with $\Web
P=\Web Q=\{1\}$) are related by a bisimulation $\cR$, then (after performing
the action $f$ on both sides), the processes $P_1\Gplus{} P_2$ and $Q_1\Gplus{}
Q_2$ (with $\Web{P_1\Gplus{} P_2}=\Web{Q_1\Gplus{} Q_2}=\{1,2\}$, and $P_i$ and
$Q_i$ located at $i$ for $i=1,2$) should be related by $\cR$. But this cannot
be achieved by saying that $P_1\Rel{\cR}Q_2$ for instance: if $P_1$ manifests
some communication capability $a$, we should insist that the same capability
$a$ be manifested by $Q_1$.

A convenient way to enforce this discipline is to say that a bisimulation is a
set of triples $(P,E,Q)$ where $P$ and $Q$ are processes and $E\subseteq\Web
P\times\Web Q$. In the example above, we start with $(P,\{(1,1)\},Q)\in\cR$
(where $1$ is the location of $\PREF f{(P_1,P_2)}$ in $P$ and similarly for
$Q$), and then, after having performed the action $f$ on both sides, we arrive
to $(P_1\Gplus{} Q_1,\{(1,1),(2,2)\},P_2\Gplus{}Q_2)\in\cR$.

Let us come back to the concept of parallel extension of a bisimulation
$\cR$. The bisimulation $\cR$ is a set of triples $(P,E,Q)$ as explained
above. We shall say that $(U,F,V)\in\cR'$ when we can find a process $S$
and two relations $C\subseteq\Web P\times\Web S$ and $D\subseteq\Web
Q\times\Web S$ with $U=P\Gplus CS$ and $V=Q\Gplus DS$. We require moreover the
existence of a relation $E$ such that $(P,E,Q)\in\cR$ and $F=E\cup\Id_{\Web S}$
(in other words, $(u,v)\in F$ if $(u,v)\in E$, or $u=v\in\Web S$), and we also
require $C$ and $D$ to be ``equivalent up to $E$'', meaning that, when
$(p,q)\in E$, we have $(p,s)\in C$ iff $(q,s)\in D$, which seems to be the
correct assumption in the proof that $\cR'$ is a bisimulation.

Bisimulations are usually defined in terms of a \emph{transition system}, a
very general and flexible concept which is essential in the study of
concurrency. Due to our more complex definition of bisimulations involving
triples $(P,E,Q)$ instead of pairs $(P,Q)$, it is not clear anymore how to use
transition systems in our framework; at least should we generalize them so as
to take localization of subprocesses into account. An abstract notion of
\emph{localized transition system} might be of general interest, but we prefer
to focus here on \TCCS{} and to define one particular localized transition
system of processes. Its states are processes. As usual in \CCS{}-like
formalisms, there are $\tau$-transitions between processes
$P\Rel{\Transtau\rho} P'$ corresponding to one internal reduction. 

The additional information $\rho$ is a function $\Web{P'}\to\Web P$ which
allows to trace the ``locative history'' of the reduction.  Labeled transition
have shape $P\Rel{\Translabs{f}{\vec L}{\lambda_1}{p}}P'$ where $p\in\Web P$,
$\vec L=(\List L1n)$ with $L_i\subseteq\Web{P'}$ and
$\lambda_1:\Web{P'}\to\Web P$ are again informations which allow to keep track
of the locative history of the reduction. These additional informations about
locations are sufficient to define an adequate notion of bisimulation.

\subsection{Localized transitions}
We define now this localized transition system\footnote{Again, we don't try to
  provide a general definition of this concept; this could be the object of
  further work}.

Let $P$ and $P'$ be processes.  
We write $P\Rel{\Translabs f{\vec L}{\lambda_1}{p}}P'$ if $p\in\Web P$,
$\Cansum{P(p)}=\PREF f{(\List P1n)}+S$ with $P'=\Subst P{\IPlus\vec P}p$,
$L_1=\Web{P_1}$,\dots, $L_n=\Web{P_n}$ and $\lambda_1:\Web{P'}\to\Web P$ is the
residual function defined by $\lambda_1(p')=p$ if
$p'\in\bigcup_{i=1}^nL_i$ and $\lambda_1(p')=p'$ otherwise\footnote{There are
  redundancies in these notations, for instance $\lambda_1$ is completely
  determined by the data $p$, $\vec L$. This redundancy will be useful in the
  sequel.}.

We write $P\Transtau{\lambda_1}P'$ if $P\Rel\Intred P'$ in the sense
of~\ref{sec:internal-reduction} and, with the notations of that section,
$\lambda_1:\Web{P'}\to\Web P$ is the residual function defined by
$\lambda_1(p')=p$ if $p'\in\bigcup_i\Web{P_i}$, $\lambda_1(p')=q$ if
$p'\in\bigcup_i\Web{Q_i}$, and $\lambda_1(p')=p'$ otherwise.

We define the reflexive-transitive closure $\Transtautr\lambda$ as follows. We
say that $P\Rel{\Transtautr\lambda}P'$ if there are $n\geq 1$, processes $\List
P1n$ and functions $\List\lambda 1{n-1}$ such that $P=P_1$, $P_n=P'$ and
$P_i\Rel{\Transtau{\lambda_i}}P_{i+1}$ for $i=1,\dots,n-1$, and
$\lambda=\lambda_1\Comp\cdots\Comp\lambda_{n-1}$.

We write \( P\Rel{\Wtranslabs fp{\vec L}{\lambda}{\lambda_1}{\lambda'}}P' \) if
there are processes $P_1$ and $P'_1$ such that
$P\Rel{\Transtautr\lambda}P_1\Rel{\Translabs f{\vec
    L}{\lambda_1}p}P'_1\Rel{\Transtautr{\lambda'}}P'$.

\subsection{Localized weak bisimilarity}
We introduce now our notion of weak bisimilarity which will be shown to imply
weak barbed congruence of processes. The definition is coalgebraic and is based
on a concept of bisimulation which, due to the importance of the graph
structure in the operational semantics of \TCCS{}, strongly uses locations.

A \emph{localized relation} (on processes) is a set
$\cR\subseteq\Processes\times\Part{\Locations^2}\times\Processes$ such that, if
$(P,E,Q)\in\cR$ then $E\subseteq\Locof P\times\Locof Q$. Such a relation $\cR$
is \emph{symmetric} if $(P,E,Q)\in\cR\Implies (Q,\Transp E,P)\in\cR$ where
$\Transp E=\{(q,p)\St(p,q)\in E\}$.

A \emph{(localized) weak bisimulation} is a symmetric localized relation such
that
\begin{itemize}
\item if $(P,E,Q)\in\cR$ and $P\Rel{\Transtau{\lambda_1}}P'$ then
  $Q\Rel{\Transtautr{\rho}}Q'$ with $(P',E',Q')\in\cR$ for some
  $E'\subseteq\Locof{P'}\times\Locof{Q'}$ such that, if $(p',q')\in E'$ then
  $(\lambda_1(p'),\rho(q'))\in E$ (this latter condition will be called
  \emph{condition on residuals})
\item if $(P,E,Q)\in\cR$ and $P\Rel{\Translabs f{\vec L}{\lambda_1}p}P'$ then
  $Q\Rel{\Wtranslabs fq{\vec M}{\rho}{\rho_1}{\rho'}}Q'$ with $(p,\rho(q))\in
  E$ and $(P',E',Q')\in\cR$ for some $E'\subseteq\Locof{P'}\times\Locof{Q'}$
  such that if $(p',q')\in E'$ then $(\lambda_1(p'),\rho\rho_1\rho'(q'))\in E$,
  and, moreover, if $n\geq 2$, then either
  $(p',\rho'(q'))\in\bigcup_{i=1}^n(L_i\times M_i)$ or
  $p'\notin\bigcup_{i=1}^nL_i$ and $\rho'(q')\notin\bigcup_{i=1}^nM_i$ (this
  condition is called \emph{condition on residuals}).
\end{itemize}
This latter dichotomy, according to whether $n=1$ or $n\geq 2$ (where $n$ is
the arity of $f$) is essential in order to obtain three effects which seem
impossible to conciliate otherwise:
\begin{itemize}
\item weak bisimilarity must be transitive
\item it must imply weak barbed congruence
\item and it should be an extension of the standard weak bisimilarity of $\CCS$
  (considering $\CCS$ as a subsystem of $\TCCS$ as explained in
  Section~\ref{sec:CCS-subsystem}).  
\end{itemize}

\begin{lemma}\label{lemma:weak-bisim-tau-iter}
  Let $\cR$ be a weak bisimulation. If $(P,E,Q)\in\cR$ and
  $P\Rel{\Transtautr\lambda}P'$, then $Q\Rel{\Transtautr\rho}Q'$ with
  $(P',E',Q')\in\cR$ for some $E'\subseteq\Locof{P'}\times\Locof{Q'}$ such that
  if $(p',q')\in E'$ then $(\lambda'(p'),\rho'(q'))\in E$.
\end{lemma}
\Beginproof
Simple induction on the length of the sequence of reductions
$P\Rel{\Transtautr\lambda}P'$.
\Endproof

\begin{lemma}\label{lemma:Transtautr-Wtranslab-comp}
  If $P\Rel{\Transtautr\lambda}P_1$, $P_1\Rel{\Wtranslabs f{p}
    {\vec L}{\lambda_1}{\lambda_2}{\lambda'_1}}P'_1$ and
  $P'_1\Rel{\Transtautr{\lambda'}}P'$ then $P\Rel{\Wtranslabs fp{\vec
      L}{\lambda\lambda_1}{\lambda_2}{\lambda'_1\lambda'}}P'$.
\end{lemma}
\Beginproof
Results immediately from the definitions.
\Endproof

Now we provide a characterization of weak bisimulation which is more symmetric
than the definition above of these relations.
\begin{lemma}\label{lemma:wbisim-charact}
  A symmetric localized relation
  $\cR\subseteq\Processes\times\Part{\Locations^2}\times\Processes$ is a weak
  bisimulation iff the following properties hold.
  \begin{itemize}
  \item If $(P,E,Q)\in\cR$ and $P\Rel{\Wtranslabs fp{\vec
        L}{\lambda}{\lambda_1}{\lambda'}}P'$, then $Q\Rel{\Wtranslabs fq{\vec
        M}{\rho}{\rho_1}{\rho'}}Q'$ with $(\lambda(p),\rho(q))\in E$ and
    $(P',E',Q')\in\cR$ for some $E'\subseteq\Locof{P'}\times\Locof{Q'}$ such
    that if $(p',q')\in E'$ then
    $(\lambda\lambda_1\lambda'(p'),\rho\rho_1\rho'(q'))\in E)$ and, moreover,
    if $n\geq 2$, either $(\lambda'(p'),\rho'(q'))\in\bigcup_{i=1}^n(L_i\times
    M_i)$ or $\lambda'(p')\notin\bigcup_{i=1}^nL_i$ and
    $\rho'(q')\notin\bigcup_{i=1}^nM_i$.
  \item If $(P,E,Q)\in\cR$ and $P\Rel{\Transtautr\lambda}P'$, then
    $Q\Rel{\Transtautr\rho}Q'$ with $(P',E',Q')\in\cR$ for some
    $E'\subseteq\Locof{P'}\times\Locof{Q'}$ such that if $(p',q')\in E'$ then
    $(\lambda(p'),\rho(q'))\in E$.
  \end{itemize}
\end{lemma}
\Beginproof
The stated property are obviously sufficient, we prove that the first one is
necessary (necessity of the second one is
Lemma~\ref{lemma:weak-bisim-tau-iter}).  Assume that $(P,E,Q)\in\cR$ and
$P\Rel{\Wtranslabs fp{\vec L}{\lambda}{\lambda_1}{\lambda'}}P'$, that is
$P\Rel{\Transtautr\lambda}P_1\Rel{\Translabs f{\vec
    L}{\lambda_1}p}P'_1\Rel{\Transtautr{\lambda'}}P'$. By
Lemma~\ref{lemma:weak-bisim-tau-iter} one has $Q\Rel{\Transtautr\rho}Q_1$ with
$(P_1,E_1,Q_1)\in\cR$ where $E_1$ is such that $(p_1,q_1)\in E_1\Implies
(\lambda(p_1),\rho(q_1))\in E$. 

Since $P_1\Rel{\Translabs f{\vec L}{\lambda_1}p}P'_1$ and
$(P_1,E_1,Q_1)\in\cR$, one has $Q_1\Rel{\Wtranslabs fq{\vec
    M}{\rho_1}{\rho_2}{\rho_1'}}{Q'_1}$ with $(p,\rho_1(q))\in E_1$ and
$(P_1',E_1',Q_1')\in\cR$ where $E'_1$ is such that if $(p'_1,q'_1)\in E'_1$
then $(\lambda_1(p'_1),\rho_1\rho_2\rho'_1(q'_1))\in E_1$ and, if $n\geq 2$,
then either $(p'_1,\rho'_1(q'_1))\in\bigcup_{i=1}^n(L_i\times M_i)$, or
$p'_1\notin\bigcup_{i=1}^nL_i$ and $\rho'_1(q'_1)\notin\bigcup_{i=1}^nM_i$.
Since $P'_1\Rel{\Transtautr{\lambda'}}P'$ and $(P_1',E_1',Q_1')\in\cR$, we can
apply Lemma~\ref{lemma:weak-bisim-tau-iter} again which shows that
$Q'_1\Rel{\Transtautr{\rho'}}Q'$ with $(P',E',Q')\in\cR$ where $E'$ is such
that $(p',q')\in E'\Implies (\lambda'(p'),\rho'(q'))\in E'_1$. By
Lemma~\ref{lemma:Transtautr-Wtranslab-comp}, we have $Q\Rel{\Wtranslabs fq{\vec
    M}{\rho\rho_1}{\rho_2}{\rho'_1\rho'}}Q'$ and remember that
$(P',E',Q')\in\cR$. We have $(p,\rho_1(q))\in E_1$ and hence
$(\lambda(p),\rho\rho_1(q))\in E$ by definition of $E_1$.  Last, the condition
on residuals obviously holds.
\Endproof

\begin{lemma}\label{lemma:weak-bisim-diag}
  Let $\cI$ be the localized relation defined by: $(P,E,Q)\in\cI$ if $P=Q$ and
  $E=\Id_{\Web P}$. Then $\cI$ is a weak bisimulation.
\end{lemma}
\Beginproof
Straightforward.
\Endproof

If $\cR$ and $\cR'$ are weak bisimulations, so is $\cR\cup\cR'$: this results
immediately from the definition. We say that $P$ and $Q$ are weakly bisimilar
(notation $P\Rel\Wbisim Q$) if there exists a weak bisimulation $\cR$ and a set
$E\subseteq\Locof P\times\Locof Q$ such that $(P,E,Q)\in\cR$.

Let $\cR$ and $\cS$ be localized relations. We define a localized relation
$\cS\Comp\cR$ as follows: $(P,H,R)\in\cS\Comp\cR$ if $H\subseteq\Locof
P\times\Locof R$ and there exist $Q$, $E$ and $F$ such that $(P,E,Q)\in\cR$,
$(Q,F,R)\in\cS$ and $F\Comp E\subseteq H$.

\begin{lemma}\label{lemma:weak-bisim-comp}
  If $\cR$ and $\cS$ are weak bisimulations, then so is $\cS\Comp\cR$.
\end{lemma}
\Beginproof
First, observe that $\cS\Comp\cR$ is symmetric.  

We use the characterization of weak bisimulations given by
Lemma~\ref{lemma:wbisim-charact}.  Let $(P,H,R)\in\cS\Comp\cR$.  Let $Q$, $E$
and $F$ be such that $(P,E,Q)\in\cR$, $(Q,F,R)\in\cS$ and $F\Comp E\subseteq
H$.

\Subcasepar Assume first that $P\Rel{\Wtranslabs fp{\vec
    L}{\lambda}{\lambda_1}{\lambda'}}P'$.  
Then we have $Q\Rel{\Wtranslabs fq{\vec M}{\rho}{\rho_1}{\rho'}}Q'$ with
$(\lambda(p),\rho(q))\in E$ and $(P',E',Q')\in\cR$ with $E'$ such that %
if $(p',q')\in E'$ %
then $(\lambda\lambda_1\lambda'(p'),\rho\rho_1\rho'(q'))\in E$ %
and, if $n\geq 2$ %
then $(\lambda'(p'),\rho'(q'))\in\bigcup_i (L_i\times M_i)$ %
or $\lambda'(p')\notin\bigcup_iL_i$ and $\rho'(q')\notin\bigcup_i M_i$. %
Therefore %
we have $R\Rel{\Wtranslabs fr{\vec N}{\sigma}{\sigma_1}{\sigma'}}R'$ %
with $(\rho(q),\sigma(r))\in F$ %
and $(Q',F',R')\in\cS$ with $F'$ such that %
if $(q',r')\in F'$ %
then $(\rho\rho_1\rho'(q'),\sigma\sigma_1\sigma'(r'))\in F)$ %
and, if $n\geq 2$ %
then $(\rho'(q'),\sigma'(r'))\in\bigcup_i (M_i\times N_i)$ %
or $\rho'(q')\notin\bigcup_i M_i$ and $\sigma'(r')\notin\bigcup_iN_i$. %
So we have $(\lambda(p),\sigma(r))\in F\Comp E\subseteq H$. 
Let
\begin{align*}
  H'&=\{(p',r')\in\Locof{P'}\times\Locof{R'}\St
  (\lambda\lambda_1\lambda'(p'),\sigma\sigma_1\sigma'(r'))\in H \text{ and if
  }n\geq 2\text{ then }
  \\
  &\quad (\lambda'(p'),\sigma'(r')) \in\bigcup_{i=1}^n (L_i\times N_i)\text{ or
  }\lambda'(p')\notin\bigcup_{i=1}^nL_i\text{ and }\sigma'(r')\notin
  \bigcup_{i=1}^nN_i\}
\end{align*}

By definition of $H'$, the triple $(P',H',R')$ satisfies the conditions on
residuals, and we are left with proving that $F'\Comp E'\subseteq H'$ which
will show that $(P',H',R')\in\cS\Comp\cR$. Let $(p',r')\in F'\Comp E'$, there
exists $q'$ such that $(p',q')\in E'$ and $(q',r')\in F'$. 

We know that $(\lambda\lambda_1\lambda'(p'),\rho\rho_1\rho'(q'))\in E$ and
$(\rho\rho_1\rho'(q'),\sigma\sigma_1\sigma'(r))\in F$ and therefore
$(\lambda\lambda_1\lambda'(p'),\sigma\sigma_1\sigma'(r))\in F\Comp E\subseteq
H$. 
So assume now that $n\geq 2$. We must prove that if
$\lambda'(p')\in\bigcup_{i=1n}L_i$ or $\sigma'(r')\in\bigcup_{i=1}^nN_i$ then
$(\lambda'(p'),\sigma'(r'))\in L_i\times N_i$ for some $i$.
Without loss of generality, we can assume that
$\lambda'(p')\in\bigcup_{i=1n}L_i$ (because the situation is symmetric).
Then by the condition on residuals for $E'$ we know that
$(\lambda'(p'),\rho'(q'))\in L_j\times M_j$ for some
$j\in\{1,\dots,n\}$, because $n\geq 2$. 
Therefore $(\rho'(q'),\sigma'(r'))\in M_i\times N_i$ by the conditions on
residuals satisfied by $F'$. It follows that $(\lambda'(p'),\sigma'(r'))\in
L_i\times N_i$ as required.

\Subcasepar Assume now that $P\Rel{\Transtautr\lambda}P'$. Since
$(P,E,Q)\in\cR$ we have $Q\Rel{\Transtautr\rho}Q'$ and there exists $E'$ such
that $(P',E',Q')\in\cR$ and, if $(p',q')\in E'$, then
$(\lambda(p'),\rho(q'))\in E$. Since $(Q,F,R)\in\cS$, we have
$R\Rel{\Transtautr\sigma}R'$ and there exists $F'$ such that $(Q',F',R')\in\cS$
and for any $(q',r')\in F'$, one has $(\rho(q'),\sigma(r'))\in F$. We have
$(P',F'\Comp E',Q')\in\cS\Comp\cR$ and it is obvious that $F'\Comp E'$
satisfies the condition on residuals.
\Endproof

We say that two processes $P$ and $Q$ are weakly bisimilar, and write
$P\Rel\Wbisim Q$, if there exists a weak bisimulation $\cR$ and a relation
$E\subseteq\Web P\times\Web Q$ such that $(P,E,Q)\in\cR$.
\begin{proposition}
  The relation $\Wbisim$ is an equivalence relation on processes.
\end{proposition}
\Beginproof
Reflexivity results from Lemma~\ref{lemma:weak-bisim-diag}, and symmetry from
the symmetry hypothesis on weak bisimulations. Transitivity is a
straightforward consequence of Lemma~\ref{lemma:weak-bisim-comp}.
\Endproof


\begin{proposition}\label{prop:wbisim-wbarbbis}
  If $P\Rel\Wbisim Q$ then $P\Rel\Wbarbbis Q$.
\end{proposition}
\Beginproof
Let $\cR$ be a weak bisimulation. Let $\cB$ be the binary relation on processes
defined by: $(P,Q)\in\cB$ if there exists $E\subseteq\Web P\times\Web Q$ such
that $(P,E,Q)\in\cR$. We contend that $\cB$ is a weak barbed bisimulation, and
this will prove the proposition. First observe that $\cB$ is symmetric because
$\cR$ is a symmetric localized relation. 

\Subcasepar
Let $(P,Q)\in\cB$ and assume first that $P\Rel\Intredtr P'$, that is
$P\Rel{\Transtautr\lambda}P'$ for some residual function $\lambda$. Let
$E\subseteq\Web P\times\Web Q$ be such that $(P,E,Q)\in\cR$. By
Lemma~\ref{lemma:wbisim-charact}, one has $Q\Rel{\Transtautr\rho}Q'$ for some
residual function $\rho$, and there exists $E'\subseteq\Web{P'}\times\Web{Q'}$
such that $(P',E',Q')\in\cR$ and therefore $(P',Q')\in\cB$ as required; this
shows that $\cB$ is a weak reduction bisimulation. 

\Subcasepar Assume now that $(P,Q)\in\cB$ and that $P\Rel\Intredtr P'$ with
$\Barb{P'}f$ (with $f\in\Barset\Sigma$ of arity $n$), meaning that
$P'\Rel{\Translabs f{\vec L}{\lambda'_1}{p'}}P''$ for some $p'\in\Locof{P'}$,
some sequence of sets of locations $\vec L$ and some residual function
$\lambda'_1$. 

Let $E\subseteq\Web P\times\Web Q$ be such that $(P,E,Q)\in\cR$. By
Lemma~\ref{lemma:wbisim-charact}, one has $Q\Rel{\Transtautr\rho}Q'$ for some
residual function $\rho$, and there exists $E'\subseteq\Web{P'}\times\Web{Q'}$
such that $(P',E',Q')\in\cR$.  Since $\cR$ is a weak bisimulation we have
therefore $Q'\Rel{\Wtranslabs f{q'}{\vec M}{\rho'}{\rho_1}{\rho''}}Q''$ and
hence $Q'\Rel\Intredtr Q_1'$ with $\Barb{Q'_1}f$. This shows that $\cB$ is
weak barb preserving since $Q\Rel\Intredtr Q_1'$.
\Endproof

We want now to prove a much stronger result, namely that weak bisimilarity
implies weak barbed congruence (and not just weak barbed bisimilarity). This
boils down to proving that weak bisimilarity is a congruence. Let us first give
an example which illustrates this implication.

\begin{example}
  Let first $\Sigma$ be such that $\Sigma_1=\{a,b\}$ and $\Sigma_i=\emptyset$
  if $i\not=1$. Then it is easy to see that $\PARFULL{\PREF a\EMPTY}{\PREF
    b\EMPTY}$ and $\PREF a{\PREF b\EMPTY}+\PREF b{\PREF a\EMPTY}$ are weakly
  bisimilar just as in usual \CCS.

  Let now $\Sigma$ be such that $\Sigma_1=\{a\}$, $\Sigma_2=\{f,g\}$ and
  $\Sigma_i=\emptyset$ for $i>2$. Let $P=\PREF f{(\PREF
    g{(\EMPTY,\EMPTY)},\EMPTY)}+\PREF g{(\PREF f{(\EMPTY,\EMPTY)},\EMPTY)}$ and
  $Q=\PARFULL{\PREF f{(\EMPTY,\EMPTY)}}{\PREF g{(\EMPTY,\EMPTY)}}$. Then we
  cannot prove that $P$ and $Q$ are weakly bisimilar (because, in the
  definition of a localized bisimulation, we are in the case $n>1$). And
  indeed, surprisingly, $P$ and $Q$ are not weak barbed bisimilar. Actually,
  let $R=\PREFO{f}{(\EMPTY,\PREFO g{(\PREF a\EMPTY,\EMPTY)}))}$. Then $\PARFULL
  QR\Rel\Intredtr\PREF a\EMPTY$ and $\Barb{\PREF a\EMPTY}a$ whereas there is no
  process $M$ such that $\PARFULL PR\Rel\Intredtr M$ with $\Barb Ma$. The best
  we can do is reduce $\PARFULL PR$ to $\PAREMPTY{\PREF
    g{(\EMPTY,\EMPTY)}}{\PREFO{g}{(\PREF a\EMPTY,\EMPTY)}}$.
\end{example}

\section{Weak bisimilarity is a congruence}
\label{sec:adequacy}

As in the standard method used in ordinary \CCS{}, the main step for proving
that weak bisimilarity is a congruence consists in extending a localized
relation $\cR$ on processes into another localized relation $\cR'$ which is,
intuitively, a congruence wrt.~``parallel composition''. Since parallel
composition here is parametrized by a relation, the definition is more involved
than in ordinary \CCS{} and strongly involves locations.

\paragraph{Adapted triples of relations.}
We say that a triple of relations $(D,D',E)$ with $D\subseteq A\times B$,
$D'\subseteq A\times B'$ and $E\subseteq B\times B'$ is \emph{adapted}, if, for
any $(a,b,b')\in A\times B\times B'$, with $(b,b')\in E$, one has $(a,b)\in D$
iff $(a,b')\in D'$.

\paragraph{Parallel extension of a localized relation.}
Let $\cR$ be a localized relation on processes. One defines a new localized
relation $\cR'$ by stipulating that $(U,F,V)\in\cR'$ if there is a process $S$,
and a triple $(P,E,Q)\in\cR$ as well are two relations $C\subseteq\Web
S\times\Web P$ and $D\subseteq\Web S\times\Web Q$ such that $U=S\Gplus CP$,
$V=S\Gplus DQ$ (these notations are introduced in Section~\ref{sec:notations}),
the triple of relations $(C,D,E)$ is adapted and $F$ is the relation $\Id_{\Web
  S}\cup E\subseteq\Web U\times\Web V$. This localized relation will be called
\emph{the parallel extension} of $\cR$.

\smallbreak

Intuitively, we express here that $U$ is the parallel composition of $S$ and
$P$, with connections between the processes of $S$ and those of $P$ specified
by $C$. And similarly for $V$, defined as the parallel composition of $S$ and
$Q$ through the relation $D$. The hypothesis that $(C,D,E)$ should be adapted
means that $C$ and $D$ specify the same connections between processes up to
$E$.

\begin{lemma}\label{lemma:graph-wbisim-sym}
  If $\cR$ is symmetric, then so is its parallel extension $\cR'$.
\end{lemma}
\Beginproof
Observe that $(C,D,E)$ is adapted iff $(D,C,\Transp E)$ is adapted.
\Endproof

The next proposition is an essential tool for proving that weak bisimulation is
a congruence.

\begin{proposition}\label{prop:parext-bisim}
  If $\cR$ is a weak bisimulation, so is its parallel extension $\cR'$.
\end{proposition}
\Beginproof
Symmetry of $\cR'$ results from the symmetry of $\cR$ and from
Lemma~\ref{lemma:graph-wbisim-sym}.  

Let $(U,F,V)\in\cR'$ with $U=S\Gplus CP$,
$V=S\Gplus DQ$, $(P,E,Q)\in\cR$, $(C,D,E)$ adapted and $F=\Id_{\Web S}\cup E$.

\paragraph{Case of a $\tau$-transition.}
Assume that $U\Rel{\Transtau\lambda}U'$. We must show that
$V\Rel{\Transtautr\rho}V'$ with $(U',F',V')\in\cR'$ and
$(\lambda(u'),\rho(v'))\in F$ for each $(u',v')\in F'$ (condition on
residuals). There are three cases as to the locations of the two guarded sums
involved in that reduction.

\Subcasepar Assume first that they are located in $S$, in other words there are
$s,t\in\Web S$ with $s\Coh St$, $\Cansum{S(s)}=\PREF f{\vec S}+\tilde S$
($\tilde S$ is a guarded sum) and $\Cansum{S(t)}=\PREFO f{\vec T}+\tilde T$
($\tilde T$ is a guarded sum), and we have $S\Transtau\mu S'$ with
\begin{itemize}
\item $\Web{S'}=(\Web
  S\setminus\{s,t\})\cup\bigcup_i\Web{S_i}\cup\bigcup_i\Web{T_i}$
\item and $\Coh{S'}$ is the least symmetric relation on $\Web{S'}$ such that
  $s'\Coh{S'}t'$ if $s'\Coh{S_i}t'$, or $s'\Coh{T_i}t'$, or
  $(s',t')\in\Web{S_i}\times\Web{T_i}$ for some $i=\{1,\dots,n\}$, or
  $\{s',t'\}\not\subseteq\bigcup_{i=1}^n\Web{S_i}\cup\bigcup_{i=1}^n\Web{T_i}$
  and $\mu(s')\Coh S\mu(t')$.
%
\end{itemize}
Remember that the residual function $\mu$ is given by $\mu(s')=s$ if
$s'\in\bigcup_i\Web{S_i}$, $\mu(s')=t$ if $s'\in\bigcup_i\Web{T_i}$ and
$\mu(s')=s'$ otherwise. We have $U'=S'\Gplus{C'}P$ where
$C'=\{(s',p)\in\Web{S'}\times\Web P\St(\mu(s'),p)\in C\}$ and
$\lambda=\mu\cup\Id_{\Web P}$.

Then we have similarly $V=S\Gplus DQ\Rel{\Transtau\rho}V'=S'\Gplus{D'}Q$ with
$\rho=\mu\cup\Id_{\Web Q}$, and $D'=\{(s',q)\in\Web{S'}\times\Web
Q\St(\mu(s'),q)\in D\}$.

The triple $(C',D',E)$ is adapted: let $s'\in\Web{S'}$, $p\in\Web P$ and
$q\in\Web Q$ be such that $(p,q)\in E$. If $(s',p)\in C'$ then $(\mu(s'),p)\in
C$ and hence $(\mu(s'),q)\in D$ since $(C,D,E)$ is adapted, that is $(s',q)\in
D'$, and similarly for the converse implication.

Coming back to the definition of $\cR'$, we see that $(U',F',V')\in\cR'$ where
$F'=\Id_{\Web{S'}}\cup E$. Moreover, the condition on residuals is satisfied,
since, given $(u',v')\in F'$, we have either $u'=v'\in\Web{S'}$ and then
$\lambda(u')=\rho(v')\in\Web S$ or $(u',v')\in E$ and
$(\lambda(u'),\rho(v'))=(u',v')\in E$. In both cases $(\lambda(u'),\rho(v'))\in
F$. 

\Subcasepar Assume next that they are located in $P$, in other words there are
$p,r\in\Web P$ with $\Cansum{P(p)}=\PREF f{\vec P}+\tilde P$ (where $\tilde P$
is a guarded sum) and $\Cansum{P(r)}=\PREFO f{\vec{R}}+\tilde{R}$ (where
$\tilde R$ is a guarded sum), and we have $P\Transtau\mu P'$ with
\begin{itemize}
\item $\Web{P'}=(\Web
  P\setminus\{p,r\})\cup\bigcup_i\Web{P_i}\cup\bigcup_i\Web{R_i}$
\item and $\Coh{P'}$ is the least symmetric relation on $\Web{P'}$ such that
  $p'\Coh{P_i}r'$ or $p'\Coh{R_i}r'$ or $(p',r')\in\Web{P_i}\times\Web{R_i}$
  for some $i\in\{1,\dots,n\}$, or
  $\{p',r'\}\not\subseteq\bigcup_i\Web{P_i}\cup\bigcup_i\Web{R_i}$ and
  $\mu(p')\Coh P\mu(r')$.
\end{itemize}
We recall that the residual function $\mu$ is given by $\mu(p')=p$ if
$p'\in\bigcup_i\Web{P_i}$, $\mu(p')=r$ if $p'\in\bigcup_i\Web{R_i}$ and
$\mu(p')=p'$ otherwise. With these notations, the process $U'$ is
$U'=S\Gplus{C'}P'$ where $C'=\{(s,p')\in\Web S\times\Web{P'}\St(s,\mu(p'))\in
C\}$ and the residual function $\lambda$ is defined as $\lambda=\Id_{\Web
  S}\cup\mu$. Since $(P,E,Q)\in\cR$ and $P\Transtautr\mu P'$, one has
$Q\Rel{\Transtautr{\nu}}Q'$ with $(P',E',Q')\in\cR$ where
$E'\subseteq\Web{P'}\times\Web{Q'}$ satisfies the condition on residuals
$(p',q')\in E'\Implies(\mu(p'),\nu(q'))\in E$. Let $D'=\{(s,q')\in\Web
S\times\Web{Q'}\St(s,\nu(q'))\in D\}$.  Setting $V'=S\Gplus{D'}Q'$, we have
$V\Rel{\Transtautr\rho}V'$ where $\rho=\Id_{\Web S}\cup\nu$.

The triple $(C',D',E')$ is adapted: let $(p',q')\in E'$ and let $s\in\Web
S$. If $(s,p')\in C'$, we have $(s,\mu(p'))\in C$. Since $(\mu(p'),\nu(q'))\in
E$ (by definition of $E'$), we have $(s,\nu(q'))\in D$ because $(C,D,E)$ is
adapted. That is $(s,q')\in D'$. The converse implication is proved similarly.

Let $F'=\Id_{\Web S}\cup E'\subseteq\Web{U'}\times\Web{V'}$, we have therefore
$(U',F',V')\in\cR'$ (by definition of $\cR'$). Last we check the condition on
residuals. Let $(u',v')\in F'$, then either $u'=v'\in\Web S$ and then
$\lambda(u')=u'=v'=\rho(v')$ or $u'\in\Web{P'}$, $v'\in\Web{Q'}$ and
$(u',v')\in E'$ and then $(\lambda(u'),\rho(v'))=(\mu(u'),\nu(v'))\in E$ by the
condition on residuals satisfied by $E$.

\Subcasepar Assume last that one of the involved guarded sums is located in $S$
and that the other one is located in $P$, this is of course the most
interesting case in this first part of the proof. 

By definition of internal reduction (see Section~\ref{sec:internal-reduction})
we have $s\in\Web S$ and $p\in\Web P$ with $(s,p)\in C$ and with
$\Cansum{S(s)}=\PREFO f{\vec S}+\tilde S$ and $\Cansum{P(p)}=\PREF f{\vec
  P}+\tilde P$ with the usual notational conventions, and $U'=S'\Gplus{C'}P'$
where $S'=\Subst{S}{\IPlus\vec S}{s}$, $P'=\Subst{P}{\IPlus\vec P}{p}$, and
$C'\subseteq\Web{S'}\times\Web{P'}$ is defined as follows: $(s',p')\in C'$ if
\begin{itemize}
\item $(s',p')\in\Web{S_i}\times\Web{P_i}$ for some $i$,
\item or $(s',p')\notin(\bigcup_i\Web{S_i})\times(\bigcup_i\Web{P_i})$ and
  $(\lambda(s'),\lambda(p'))\in C$,
\end{itemize}
where the residual map $\lambda:\Web{U'}=\Web{S'}\cup\Web{P'}\to\Web{U}=\Web
S\cup\Web P$ is defined by $\lambda(u')=u'$ if
$u'\in(\Web{S'}\setminus\bigcup_i\Web{S_i})
\cup(\Web{P'}\setminus\bigcup_i\Web{P_i})$, $\lambda(s')=s$ if
$s'\in\bigcup_i\Web{S_i}$ and $\lambda(p')=p$ if $p'\in\bigcup_i\Web{P_i}$.


We have $P\Rel{\Translabs f{\vec L}\lambda p}P'$ (where $L_i=\Web{P_i}$ for
each $i=1,\dots,n$) and hence, %
since we have assumed that $(P,E,Q)\in\cR$, %
we have $Q\Rel{\Wtranslabs fq{\vec M}\rho{\rho_1}{\rho'}}Q'$ %
with $(p,\rho(q))\in E$ %
and $(P',E',Q')\in\cR$ where $E'$ is such that if $(p',q')\in E'$ then %
$(\lambda(p'),\rho\rho_1\rho'(q'))\in E$ %
and, if $n\geq 2$, then %
 $(p',\rho'(q'))\in L_i\times M_i$ for some $i$, or
$p'\notin\bigcup_i L_i$ and $\rho'(q')\notin\bigcup M_i$. 

We can decompose this transition as follows
$$
Q\Rel{\Transtautr\rho}Q_1\Rel{\Translabs f{\vec
    M}{\rho_1}q}Q'_1\Rel{\Transtautr{\rho'}}Q'\,.
$$
With these notations we have $V\Rel{\Transtautr{\mu}}V_1$ with
$V_1=S\Gplus{D_1}Q_1$ where $D_1=\{(s,q_1)\in\Web S\times\Web{Q_1}\St
(s,\rho(q_1))\in D\}$, and $\mu=\Id_{\Web S}\cup\rho$.

We have $q\in\Web{Q_1}$ with $\Cansum{Q_1(q)}=\PREF f{\vec R}+\tilde R$ and
$\Web{R_i}=M_i$ for $i=1,\dots,n$. Moreover, since $(p,\rho(q))\in E$ and
$(s,p)\in C$, and since $(C,D,E)$ is adapted, we have $(s,\rho(q))\in D$, that
is $(s,q)\in D_1$. Therefore, since $\Cansum{S(s)}=\PREFO f{\vec S}+\tilde S$,
we have $V_1\Rel{\Transtau\theta}V'_1=S'\Gplus{D'_1}Q'_1$ where
$D'_1\subseteq\Web{S'}\times\Web{Q'_1}$ is defined as follows: given
$(s',q'_1)\in\Web{S'}\times\Web{Q'_1}$, we have $(s',q'_1)\in D'_1$

\begin{itemize}
\item if $s'\in\Web{S_i}$ and $q'_1\in\Web{R_i}$ for some $i=1,\dots,n$
\item or $s'\notin\bigcup_i\Web{S_i}$ or $q'_1\notin\bigcup_i\Web{R_i}$ and
  $(\theta(s'),\theta(q'_1))\in D_1$ (that is $(\theta(s'),\rho\theta(q'_1))\in
  D$),
\end{itemize}
and the residual function $\theta$ is defined by %
$\theta(v'_1)=v'_1$ if %
$v'_1\in(\Web
S\setminus\bigcup_i\Web{S_i})\cup(\Web{Q_1}\setminus\bigcup_i\Web{R_i})$, %
$\theta(s')=s$ if $s'\in\bigcup_i\Web{S_i}$ and %
$\theta(q'_1)=q$ if $q'_1\in\bigcup_i\Web{R_i}$. 

Observe that $\theta(q'_1)=\rho_1(q'_1)$ for all $q'_1\in\Web{Q'_1}$.

Since $Q'_1\Rel{\Transtautr{\rho'}}Q'$, we have
$V'_1=S'\Gplus{D'_1}Q'_1\Rel{\Transtautr{\mu'}}V'=S'\Gplus{D'}Q'$ where
$\mu'=\Id_{\Web{S'}}\cup\rho'$ and
$D'=\{(s',q')\in\Web{S'}\times\Web{Q'}\St(s',\rho'(q'))\in D'_1\}$. So we have
$V\Rel{\Transtautr{\mu\theta\mu'}}V'$. Let
$F'\subseteq\Web{U'}\times\Web{V'}$ be defined by
$F'=\Id_{\Web{S'}}\cup E'$. It is clear then that $(u',v')\in
F'\Implies(\lambda(u'),\mu\theta\mu'(v'))\in F$ because $(p',q')\in
E'\Implies(\lambda(p'),\rho\rho_1\rho'(q'))\in E$ and $\theta$ and $\rho_1$
coincide on $\Web{Q'_1}$.

To finish, we must prove that $(U',F',V')\in\cR'$ and to this end it suffices
to show that the triple of relations $(C',D',E')$ is adapted. So let
$s'\in\Web{S'}$, $p'\in\Web{P'}$ and $q'\in\Web{Q'}$ with $(p',q')\in
E'$ (so that in particular $(\lambda(p'),\rho\theta\rho'(q'))\in E$). 

Assume first that $(s',p')\in C'$ and let us show that $(s',q')\in D'$, that is
$(s',\rho'(q'))\in D'_1$. Coming back to the definition of $C'$, we can reduce
our analysis to three cases.
\begin{itemize}
\item First case: $(s',p')\in\Web{S_i}\times\Web{P_i}$ for some $i$. We
  distinguish two cases as to the value of $n$ (the arity of $f$). Assume first
  that $n\geq 2$. Since $p'\in\Web{P_i}=L_i$, we must have $\rho'(q')\in
  M_i=\Web{Q_i}$ because $(p',q')\in E'$ and then $(s',\rho'(q'))\in D'_1$ as
  required. Assume now $n=1$. If $\rho'(q')\in M_1$ we reason as above, so
  assume that $\rho'(q')\notin M_1=\bigcup_{i=1}^n\Web{R_i}$. Coming back to
  the definition of $D'_1$, it suffices to prove that
  $(\theta(s'),\rho\theta\rho'(q'))=(s,\rho\rho'(q'))\in D$. Since $(p',q')\in
  E'$ we have $(\lambda(p'),\rho\theta\rho'(q'))=(p,\rho\rho'(q'))\in E$. We
  also have $(s,p)\in C$, and hence $(s,\rho\rho'(q'))\in D$ as required, since
  $(C,D,E)$ is adapted.
\item Second case: $s'\notin\bigcup_i\Web{S_i}$. In order to prove $(s',q')\in
  D'$, it suffices to prove that
  $(\theta(s'),\rho\theta\rho'(q'))=(s',\rho\theta\rho'(q'))\in D$. But we have
  $(s',p')\in C'$ and $s'\notin\bigcup_i\Web{S_i}$, hence
  $(\lambda(s'),\lambda(p'))=(s',\lambda(p'))\in C$. Since $(p',q')\in E'$, we
  have $(\lambda(p'),\rho\theta\rho'(q'))\in E$ and hence
  $(s',\rho\theta\rho'(q'))\in D$ since $(C,D,E)$ is adapted.
\item Third case: $s'\in\bigcup_i\Web{S_i}$ and $p'\notin\bigcup_i\Web{P_i}$ so
  that we have $(s,p')\in C$ (by definition of $C'$ and because $(s',p')\in
  C'$). Assume first that $n\geq 2$. Since $(p',q')\in E'$, we must have
  $\rho'(q')\notin\bigcup_{i=1}^nM_i$. To prove that $(s',\rho'(q'))\in D'_1$,
  it suffices therefore to check that
  $(\theta(s'),\rho\theta\rho'(q'))=(s,\rho\rho'(q'))\in D$. This property
  holds because $(C,D,E)$ is adapted, $(s,p')\in C$ and $(p',\rho\rho'(q'))\in
  E$ because $(p',q')\in E'$. Assume now that $n=1$. If
  $\rho'(q')\notin\bigcup_{i=1}^nM_i=M_1$, we can reason as above, so assume
  that $\rho'(q')\in M_1$. Then we have $(s',\rho'(q'))\in\Web{S_1}\times M_1$
  and hence $(s',\rho'(q'))\in D'_1$.
\end{itemize}
    
Let us prove now the converse implication, assuming that $(s',\rho'(q'))\in
D'_1$; we contend that $(s',p')\in C'$. Again, we consider three cases.
\begin{itemize}
\item First case: $s'\in\Web{S_i}$ and $\rho'(q')\in M_i=\Web{R_i}$
  for some $i\in\{1,\dots,n\}$. If $n\geq 2$ the fact that $(p',q')\in E'$
  implies that $p'\in L_i=\Web{P_i}$ and hence $(s',p')\in C'$ as
  required. Assume that $n=1$ and $p'\notin L_1=\bigcup_{i=1}^n\Web{P_i}$, we
  have $(\lambda(s'),\lambda(p'))=(s,p')\in C$ because $(s,\rho\rho'(q'))\in D$
  ---~since $(s',\rho'(q'))\in D'_1$, $\rho'(q')\notin M_1$ and
  $(\theta(s'),\rho\theta\rho'(q'))=(s,\rho\rho'(q'))$~---,
  $(p,\rho\rho'(q'))\in E$ and $(C,D,E)$ is adapted. Hence $(s',p')\in C'$.
\item Second case: $s'\notin\bigcup_i\Web{S_i}$. In view of the definition
  of $C'$, it suffices to prove that
  $(\lambda(s'),\lambda(p'))=(s',\lambda(p'))\in C$. Since $(s',\rho'(q'))\in
  D'_1$ and $s'\notin\bigcup_i\Web{S_i}$, we have
  $(\theta(s'),\rho\theta\rho'(q'))=(s',\rho\theta\rho'(q'))\in D$. And since
  $(p',q')\in E'$ we have $(\lambda(p'),\rho\theta\rho'(q'))\in E$, and hence
  $(s',\lambda(p'))\in C$ because $(C,D,E)$ is adapted.
\item Third case: $s'\in\Web{S_i}$ for some $i\in\{1,\dots,n\}$ and
  $\rho'(q')\notin\bigcup_iM_i$. If $n\geq 2$, we must have
  $p'\notin\bigcup_iL_i$ because $(p',q')\in E'$. Therefore, to check that
  $(s',p')\in C'$, it suffices to prove that
  $(\lambda(s'),\lambda(p'))=(s,p')\in C$. We have $(s',\rho'(q'))\in D'_1$ and
  hence $(\theta(s'),\rho\theta\rho'(q'))=(s,\rho\rho'(q'))\in D$. Since
  $(p',q')\in E'$ we have
  $(\lambda(p'),\rho\theta\rho'(q'))=(p',\rho\rho'(q'))\in E$ and hence
  $(s,p')\in C$ because $(C,D,E)$ is adapted. Assume now that $n=1$. If $p'\in
  L_1$ we have $(s',p')\in C'$ since $(s',p')\in\Web{S_1}\times\Web{P_1}$. So
  assume that $p'\notin L_1$. Since then $p'\notin\bigcup_{i=1}^n\Web{P_i}$, it
  suffices to prove that $(\lambda(s'),\lambda(p'))=(s,p')\in C$ (by definition
  of $C'$). We have $(p',\rho\theta\rho'(q'))=(p',\rho\rho'(q'))\in E$ because
  $(p',q')\in E'$ and $(s,\rho\theta\rho'(q'))=(s,\rho\rho'(q'))\in D$ because
  $(s',\rho'(q'))\in D'_1$ and $\rho'(q')\notin\bigcup_iM_i$. It follows that
  $(s,p')\in C$ as required.
\end{itemize}
 
This ends the first part of the proof.

\paragraph{Case of a labeled transition.}
We assume now that $U\Rel{\Translabs f{\vec L}{\mu_1}r}U'$. Since $U=S\Gplus
CP$, we consider two cases as to the location of $r$.

\Subcasepar If $r\in\Web S$ then we have $\Cansum{S(r)}=\PREF f{\vec S}+\tilde
S$ and $S\Rel{\Translabs f{\vec L}{\sigma_1}r}S'$ where $S'=\Subst S{\IPlus\vec
  S}r$ (so that $L_i=\Web{S_i}$ for each $i$), and $U'=S'\Gplus{C'}P$ where
$C'=\{(s',p)\in\Web{S'}\times\Web{P}\St(\sigma_1(s'),p)\in C\}$. Let
$D'=\{(s',q)\in\Web{S'}\times\Web{Q}\St(\sigma_1(s'),q)\in D\}$. We have
$\mu_1=\sigma_1\cup\Id_{\Web P}$. It is clear that $(C',D',E)$ is adapted,
since $(C,D,E)$ is adapted.

Let $V'=S'\Gplus{D'}Q$, we have just seen that $(U',F',V')\in\cR'$ where
$F'=\Id_{\Web{S'}}\cup E$. We have $(r,r)\in F$, $V\Rel{\Translabs f{\vec
    L}{\nu_1}r}V'$ (with $\nu_1=\sigma_1\cup\Id_{\Web Q}$) and, given
$(u',v')\in F'$, we have either $(u',v')\in\bigcup_i(L_i\times L_i)$ (and
actually $u'=v'$) or $u'\notin\bigcup_iL_i$, $v'\notin\bigcup_iL_i$ and
$(u',v')\in F$ as easily checked. Therefore the condition on residuals is
satisfied.

\Subcasepar The last case to consider is when $r=p\in\Web P$ and then we have
$P(p)=\PREF f{\vec P}+\tilde P$ and $P\Rel{\Translabs f{\vec
    L}{\lambda_1}p}P'$. Then we have $U'=S\Gplus{C'}P'$ where
$C'=\{(s,p')\in\Web S\times\Web{P'}\St(s,\lambda_1(p'))\in C\}$.

Since $(P,E,Q)\in\cR$ we have $Q\Rel{\Wtranslabs fq{\vec
    M}\rho{\rho_1}{\rho'}}Q'$ with $(p,\rho(q))\in E$ and there exists
$E'\subseteq\Web{P'}\times\Web{Q'}$ such that $(P',E',Q')\in\cR$ and, for any
$(p',q')\in E'$, $(\lambda_1(p'),\rho\rho_1\rho'(q'))\in E$ and, if $n\geq2$,
either $(p',\rho'(q'))\in\bigcup_{i=1}^n(L_i\times M_i)$, or
$p'\notin\bigcup_{i=1}^nL_i$ and $\rho'(q')\notin\bigcup_{i=1}^nM_i$.

Therefore we have $V\Rel{\Wtranslabs fq{\vec M}\nu{\nu_1}{\nu'}}V'$ where
$V'=S\Gplus{D'}Q'$ with $D'=\{(s,q')\in\Web
S\times\Web{Q'}\St(s,\rho\rho_1\rho'(q'))\in D\}$. Moreover $\nu=\Id_{\Web
  S}\cup\rho$, $\nu_1=\Id_{\Web S}\cup\rho_1$ and $\nu'=\Id_{\Web S}\cup\rho'$.

Let $F'\subseteq\Web{U'}\times\Web{V'}$ be defined by $F'=\Id_{\Web S}\cup
E'$. Let $(u',v')\in F'$. If $u'\in\Web S$ or $v'\in\Web S$, we must have
$u'=v'$. If $u'\notin\Web S$ and $v'\notin\Web S$ then we have $(u',v')\in E'$
and hence $(\mu_1(u'),\nu\nu_1\nu'(v'))=(\lambda_1(u'),\rho\rho_1\rho'(q'))\in
E$ and, if $n\geq 2$, either there exists $i$ such that $u'\in L_i$ and
$\nu'(v')=\rho'(v')\in M_i$, or $u'\notin\bigcup_iL_i$ and
$\nu'(v')=\rho'(v')\notin\bigcup_iM_i$.

Moreover, the triple $(C',D',E')$ is adapted: let $(p',q')\in E'$ and $s\in\Web
S$. We have $(\lambda_1(p'),\rho\rho_1\rho'(q'))\in E$. We have $(s,p')\in C'$
iff $(s,\lambda_1(p'))\in C$ iff $(s,\rho\rho_1\rho'(q'))\in D$ iff $(s,q')\in
D'$. 
\Endproof

Now we are in position of proving that weak bisimilarity is a congruence, a
result which is interesting \emph{per se} and will be essential for proving
Theorem~\ref{th:wbisim-impl-barbed-congr}.
\begin{theorem}\label{th:wbisim-th}
  The weak bisimilarity relation $\Wbisim$ is a congruence.
\end{theorem}
\Beginproof
Let $\cR$ be a weak bisimulation. Let $R$ be a $Y$-context. We define a new
localized relation denoted as $\Subst R\cR Y$:
\begin{itemize}
\item if $R=Y$ then $\Subst R\cR Y=\cR$;
\item if $R\not=Y$ then we stipulate that $(P',E',Q')\in\Subst R\cR Y$ if there
  exists $(P,E,Q)\in\cR$ and if $E'=\Id_{\Web R}$, $P'=\Subst RPY$ and
  $Q'=\Subst RQY$ (observe that $\Web{P'}=\Web{Q'}=\Web R$ because $R\not=Y$).
\end{itemize}
We define a localized relation $\Contrel\cR$ as the union of $\cI$ (the set of
all triples $(U,E,U)$ where $U$ is any process and $E=\Id_{\Web U}$), of the
parallel extension $\cR'$ of $\cR$ (see Proposition~\ref{prop:parext-bisim})
and of all the relations of the shape $\Subst R\cR Y$ for all $Y$-contexts $R$.

We prove that $\Contrel\cR$ is a weak bisimulation and the theorem will follow
easily.

\smallbreak

Let $(U,F,V)\in\Contrel\cR$ and assume that we are in one of the two following
situations
\begin{itemize}
\item $U\Rel{\Transtau\mu}U'$ (called case \Casename 1 in the sequel)
\item or $U\Rel{\Translabs f{\vec L}{\mu_1}p}U'$ (called case~\Casename 2
  in the sequel).
\end{itemize}

We describe explicitely our objectives.
\begin{itemize}
\item In case \Casename 1 we must show that $V\Rel{\Transtautr\nu}V'$ with
  $(U',F',V')\in\Contrel\cR$ for some $F'\subseteq\Web{U'}\times\Web{V'}$ such
  that for any $(u',v')\in F'$, one has $(\mu(u'),\nu(v'))\in F$.
\item In case \Casename 2 we must show that $V\Rel{\Wtranslabs fq{\vec
      M}\nu{\nu_1}{\nu'}}V'$ with $(p,\nu(q)))\in F$ and
  $(U',F',V')\in\Contrel\cR$, for some $F'\subseteq\Web{U'}\times\Web{V'}$ such
  that, for any $(u',v')\in F'$, one has $(\mu_1(u'),\nu\nu_1\nu'(v'))\in F$
  and, if $n\geq 2$, then one has either
  $(u',\nu'(v'))\in\bigcup_{i=1}^n(L_i\times M_i)$ or $u'\notin\bigcup_iL_i$ and
  $\nu'(v')\notin\bigcup_iM_i$.
\end{itemize}
 
The case where $(U,F,V)\in\cI$ is trivial.

\smallbreak

If $(U,F,V)\in\cR'$ we apply directly Proposition~\ref{prop:parext-bisim} in
both cases.

\smallbreak

Assume now that $(U,F,V)\in\Subst R\cR Y$ for some $Y$-context $R$, so that
$U=\Subst RPY$, $V=\Subst RQY$ with $(P,E,Q)\in\cR$ and $F=E$ if $R=Y$ and
$F=\Id_{\Web R}$ otherwise.  If $R=Y$ we use directly the fact that $\cR$ is a
weak bisimulation to exhibit $V'$ and $F'$ satisfying the required
conditions. 

So we assume from now on that $R\not=Y$ and therefore $F=\Id_{\Web R}$.

\smallbreak

By definition of a $Y$-context, there is exactly one $r\in\Web R$ such that $Y$
occurs free in $R(r)$. Then $R(r)$ can be written uniquely as $R(r)=\PREF
g{\vec R}+\tilde R$ where $Y$ does not occur in $\tilde R$ and occurs in
exactly one of the processes $\List R1n$; without loss of generality
we can assume that $R_1$ is a $Y$-context and that $Y$ does not occur free in
$\List R2n$.

Assume first that $R_1\not=Y$. In both cases \Casename 1 and \Casename 2, we
have $U'=\Subst{R'}PY$ with $R\Rel{\Transtau\mu}R'$ (case \Casename 1) or
$R\Rel{\Translabs f{\vec L}{\mu_1}p}R'$ (case \Casename 2). Let
$V'=\Subst{R'}{Q}Y$. In case \Casename 1, we have $V\Rel{\Transtau\mu}V'$ and
in case \Casename 2 we have $V\Rel{\Translabs f{\vec L}{\mu_1}q}V'$, and since
$R'\not=Y$ (by our hypothesis on $R_1$), we have
$(U',\Id_{\Web{R'}},V')\in\Contrel\cR$ because $(P,E,Q)\in\cR$. The condition
on residuals is obviously satisfied in both cases.

Assume now that $R_1=Y$. 

\Subcasepar Suppose first that we are in case \Casename 1. There are two cases
to consider as to the locations $s,t\in\Web U$ of the sub-processes involved in
the transition $U\Rel{\Transtau\mu}U'$. The case where $s\not=r$ and $t\not=r$
is similar to the case above where $R_1\not=Y$. By symmetry we are left with
the case where $s=r$ (and hence $t\not=r$). 

So $U(t)=R(t)=\PREFO f{\vec T}+\tilde T$ and the guarded sum $R(r)$ has an
unique summand which is involved in the transition $U\Rel{\Transtau\mu}U'$
(called \emph{active summand} in the sequel), and this summand is of the shape
$\PREF f{\vec S}$. 

If the active summand is $\PREF g{\vec R}$\footnote{Remember that $\PREF g{\vec
    R}$ is the unique summand of $R(r)$ which contains $Y$.} (so that $g=f$)
then $U(r)=\PREF f{(P,\List R2n)}+\tilde S$ and $U'$ can be written
$U'=R'\Gplus CP$ for some process $R'$ which can be defined using only $R$, and
$C\subseteq\Web{R'}\times\Web P$. Explicitly, $R'$ is defined as follows:
\begin{itemize}
\item $\Web{R'}=(\Web R\setminus\{r,t\})\cup
  \bigcup_{i=2}^n\Web{R_i}\cup\bigcup_{i=1}^n\Web{T_i}$
\item and $\Coh{R'}$ is the least symmetric relation on $\Web{R'}$ such that
  $r'\Coh{R'}t'$ if $r'\Coh{R_i}t'$ for some $i=2,\dots,n$ or $r'\Coh{T_i}t'$
  for some $i=1,\dots,n$, or $(r',t')\in\Web{R_i}\times\Web{T_i}$ for some
  $i\in\{2,\dots,n\}$, or $r'\notin\bigcup_{i=2}^n\Web{R_i}$ or
  $t'\notin\bigcup_{i=1}^n\Web{T_i}$ and $r'\Coh{R}t$ and
  $\mu(r')\Coh R\mu(t')$
\end{itemize}
where the residual function $\mu:\Web{U'}\to\Web U$ is given by $\mu(r')=r$ if
$r'\in\Web P\cup\bigcup_{i=2}^n\Web{R_i}$, $\mu(r')=t$ if
$r'\in\bigcup_{i=1}^n\Web{T_i}$ and $\mu(r')=r'$ when $r'$ belongs to none of
these two sets. 

The relation $C$ is defined as follows: given $(r',p)\in\Web{R'}\times\Web P$,
one has $(r',p)\in C$ if $r'\in\Web{T_1}$, or
$r'\notin\bigcup_{i=2}^n\Web{R_i}\cup\bigcup_{i=1}^n\Web{T_i}$ and $r'\Coh
Rr$.

Let $V'=R'\Gplus DQ$, where $D\subseteq\Web{R'}\times\Web Q$ is defined exactly
like $C$ (just replace $P$ by $Q$ in the definition). Then $(C,D,E)$ is adapted
(because the property for $(r',p)\in\Web{R'}\times\Web P$ of belonging or not
to $C$ depends only on $r'$, and does not depend on $p$, and similarly for
$D$). We can mimic that reduction on $V$, so that $V\Rel{\Transtau\nu}V'$ for
the residual function $\nu$ which is defined like $\mu$ (replacing $P$ by
$Q$). We have $(U',F',V')\in\cR'\subseteq\Contrel\cR$ where
$F'=\Id_{\Web{R'}}\cup E$. Given $(u',v')\in F'$, we have $\mu(u')=\nu(v')$,
that is $(\mu(u'),\nu(v'))\in F$ so that the condition on residuals
holds\footnote{It is in this part of the proof that one understand the
  importance of adapted triples of relations in the definition of the parallel
  extension of a weak bisimulation.}.

Assume now that the active summand is not $\PREF g{\vec R}$. In that case we
also have $V\Rel{\Transtau\mu}U'$ (both $P$ and $Q$ vanish in the corresponding
reductions), and we are done because
$(U',\Id_{\Web{U'}},U')\in\cI\subseteq\Contrel\cR$.

\Subcasepar We suppose now that we are in case \Casename 2.  Assume first that
$p\not=r$. In that case we have $R\Rel{\Translabs f{\vec L}{\theta_1}p}R'$ and
$U'=\Subst{R'}{P}{Y}$ and we also have $V\Rel{\Translabs f{\vec
    L}{\theta_1}p}V'=\Subst{R'}{Q}{Y}$ so
$(U',\Id_{\Web{R'}},V')\in\Subst{R'}{\cR}{Y}\subseteq\Contrel\cR$, and the
condition on residuals is obvious.

Assume now that $p=r$. Then exactly one of the summands of the guarded sum
$R(r)$ is the prefixed process performing the action $f$ in the considered
transition on $U$ (again, this summand is called the active summand in the
sequel).

The case where the active summand is not $\PREF g{(P,\List R2n)}$ is completely
similar to the previous one ($P$ vanishes in the transition).

Assume that the active summand is $\PREF g{(P,\List R2n)}$ (so that $g=f$),
then $U'=R'\Gplus CP$ where $R'$ is defined by
\begin{itemize}
\item $\Web{R'}=(\Web R\setminus\{r\})\cup\bigcup_{i=2}^n\Web{R_i}$ and
  $\Coh{R'}$ is the least symmetric relation on $\Web{R'}$ such that
  $r'\Coh{R'}t'$ if $r'\Coh{R_i}t'$ for some $i=2,\dots,n$ or
  $\theta_1(r')\Coh R\theta_1(t')$.
\item The relation $C\subseteq\Web{R'}\times\Web P$ is defined by $(r',q)\in C$
  if $r'\notin\bigcup_{i=2}^n\Web{R_i}$ and $r'\Coh Rr$ (this does not depend
  on $q$).
\end{itemize}
Then we have $V\Rel{\Translabs f{\vec M}{\phi_1}p}V'$ (with $M_1=\Web Q$ and
$M_i=L_i=\Web{R_i}$ for $i=2,\dots,n$) with $V'=R'\Gplus DQ$ where $D$ is
defined like $C$ (replacing $P$ by $Q$ in the definition). Then we have
$(U',F',V')\in\cR'\subseteq\Contrel\cR$ where $F'=\Id_{\Web{R'}}\cup E$ since
$(C,D,E)$ is obviously adapted (as above). Moreover the condition on residuals
is obviously satisfied. This ends the proof of the fact that $\Contrel\cR$ is a
weak bisimulation.

We can now prove that $\Wbisim$ is a congruence. Assume that $P\Rel\Wbisim Q$
and let $R$ be a $Y$-context. Let $E\subseteq\Web P\times\Web Q$ and let $\cR$
be a weak bisimulation such that $(P,E,Q)\in\cR$. Then we have $(\Subst
RPY,\Id_{\Web R},\Subst RQY)\in\Subst R\cR Y\subseteq\Contrel\cR$ and hence
$\Subst RPY\Wbisim\Subst RQY$ since $\Contrel\cR$ is a weak bisimulation.
\Endproof

We can prove now the main theorem of the paper.
\begin{theorem}\label{th:wbisim-impl-barbed-congr}
  Let $P$ and $Q$ be processes. If $P\Rel\Wbisim Q$ ($P$ and $Q$ are weakly
  bisimilar) then $P\Rel\Wbarbcong Q$ ($P$ and $Q$ are weakly barb congruent).
\end{theorem}
\Beginproof
Assume that $P\Rel\Wbisim Q$ and let $R$ be a $Y$-context. We have $\Subst
RPY\Rel\Wbisim\Subst RQY$ by Theorem~\ref{th:wbisim-th} and hence $\Subst
RPY\Rel\Wbarbbis\Subst RQY$ by Proposition~\ref{prop:wbisim-wbarbbis}. 
\Endproof

\section{Weak bisimilarity on \CCS{}}
\label{sec:CCS-bisimilarity}
We assume in this section that $\Sigma_n=\emptyset$ if $n\not=1$ (see the end
of Section~\ref{sec:automata-as-processes}).  All processes $P$ considered in
this section are \CCS{} processes built on $\Sigma$, meaning that, in any
subprocess of $P$ which is of shape $\PAR
G\Phi$, the graph $G$ is a complete graph (for all $p,q\in\Web G$, $p\Coh Gq$).
%

We answer here a very natural question: when restricted to ordinary \CCS{},
does our weak localized bisimilarity coincide with standard weak bisimilarity?

Let $\cR$ be a localised weak bisimulation. Let $\CCSbisim\cR$ be the following
relation on \CCS{} processes:
$P\Rel{\CCSbisim\cR}Q$ if $(P,E,Q)\in\cR$ for some $E\subseteq\Web P\times\Web
Q$. We prove that $\CCSbisim\cR$ is a weak bisimulation on \CCS{} processes.

\begin{lemma}
  Let $\cR$ be a localized weak bisimulation. Then $\CCSbisim\cR$ is weak
  bisimulation on $\CCS$ processes.
\end{lemma}
\Beginproof
Let $P$ and $Q$ be $\CCS$ processes such that $P\Rel{\cR^0}Q$. Let
$E\subseteq\Web P\times\Web Q$ be such that $(P,E,Q)\in\cR$.

Assume first that $P\Rel\CCStau P'$. Let $p_1,p_2\in\Web P$ with
$\Cansum{P(p_1)}=\PREFI a{P_1}+S_1$ and $\Cansum{P(p_2)}=\PREFO a{P_2}+S_2$
(the two sub-processes involved in this reduction). Then, by definition of the
internal reduction in $\TCCS$, $P'=\PAR
G\Phi$ where $G$ is the complete graph on $\Web G=\Web
P\setminus\{p_1,p_2\}\cup\Web{P_1}\cup\Web{P_2}$ and $\Phi(r)=P(r)$ if
$r\in\Web P$, $\Phi(r)=P_i(r)$ if $r\in\Web{P_i}$ for $i=1,2$. In other words
$P'=\Substbis{P}{P_1/p_1,P_2/p_2}$

Let $\lambda_1:\Web{P'}\to\Web P$ be the corresponding residual map
($\lambda_1(r)=r$ if $r\in\Web P$ and $\lambda_1(r)=p_i$ if $r\in\Web{P_i}$),
we have $P\Rel{\Transtau{\lambda_1}}P'$ and therefore there is a \TCCS{}
process $Q'$ such that $(P',E',Q')\in\cR$ for some relation
$E'\subseteq\Web{P'}\times\Web{Q'}$, and a function $\rho:\Web{Q'}\to\Web Q$
with $Q\Rel{\Transtautr\rho}Q'$ and $(p',q')\in E'\Implies
(\lambda_1(p'),\rho(q'))\in E$. Therefore we have $P'\Rel{\CCSbisim\cR}Q'$ as
required.

Assume now that $P\Rel{\CCSlab a}P'$. Let $p\in\Web P$ with
$\Cansum{P(p)}=\PREFI a{P_1}+S_1$ and $P'=\Subst P{P_1}p$. Then we have
$P\Rel{\Translabs aL{\lambda_1}p}P'$ where $L=\Web{P_1}$ and
$\lambda_1:\Web{P'}\to\Web P$ is given by $\lambda_1(r)=p$ if $r\in\Web{P_1}$
and $\lambda_1(r)=r$ otherwise. Since $(P,E,Q)\in\cR$, we have
$Q\Rel{\Wtranslabs aq{M}{\rho}{\rho_1}{\rho'}}Q'$ with $(p,\rho(q))\in E$, and
there exists $E'\subseteq\Web{P'}\times\Web{Q'}$ such that $(P',E',Q')\in\cR$,
and $(\lambda_1(p'),\rho\rho_1\rho'(q'))\in E$ for each $(p',q')\in E'$. In
particular $P'\Rel{\cR^0}Q'$.

Since $\cR$ is a localized bisimulation, the relation $\cR^0$ is symmetric and
is therefore a bisimulation on $\CCS$ processes.
\Endproof

We need now to prove the converse. Let $\cU$ be a binary relation
on $\CCS$ processes. Let $\CCSlocrel\cU$ be the set of all triples $(P,E,Q)$
where $P$ and $Q$ are $\CCS$ processes such that $P\Rel\cU Q$ and $E=\Web
P\times\Web Q$.

\begin{lemma}
  If $\cU$ is a bisimulation, then $\CCSlocrel\cU$ is a localized bisimulation.
\end{lemma}
\Beginproof
Let $P$ and $Q$ be $\CCS$ processes and let $E$ be such that
$(P,E,Q)\in\CCSlocrel\cU$, so that $E=\Web P\times\Web Q$ and $P\Rel\cU Q$.

Assume first that $P\Rel{\Transtau{\lambda_1}}P'$ so that $P\Rel{\CCStau}P'$
(in $\CCS$) and hence there exists $Q'$ such that $Q\Rel{\CCStautr}Q'$ and
$P'\Rel\cU Q'$. Then there is a function $\rho:\Web{Q'}\to\Web Q$ such that
$Q\Rel{\Transtautr{\rho}}Q'$ and we have $(P',E',Q')\in\CCSlocrel\cU$. The
condition on residuals holds obviously, by definition of $E$.

The case of a labeled transition is completely similar and the condition on
residuals holds again by definition of $\CCSlocrel\cU$ and because we are in
the case where $n=1$ (all function symbols are of arity $1$).
\Endproof

So we can conclude that, when restricted to $\CCS$ processes, our notion of
weak bisimilarity coincides with the usual one.
\begin{proposition}
  Two $\CCS$ processes are weakly bisimilar (in the usual $\CCS$ sense) iff
  they are weakly bisimilar in the localized sense.
\end{proposition}

\section*{Conclusion}
We have presented an extension of \CCS{} which deals with trees instead of
words, and various concepts and tools associated with this new process
algebra. The notion of barbed bisimilarity, as it is defined here, is a
straightforward generalization of the corresponding notion for \CCS{} and
therefore is hardly questionable, but we cannot say the same of weak
bisimilarity. It will be crucial to understand if weak bisimilarity is
equivalent to weak barbed congruence here and, if not, to look for a more
liberal notion of weak bisimilarity in order to get such a full abstraction
property. Another more conceptual task will be to extend this approach to more
expressive settings such as for instance the $\pi$-calculus, and of course to
understand if \TCCS{} can be encoded in such settings.

This work also originated from the encodings of the $\pi$-calculus and of the
solos calculus in differential interaction nets by the first author and
Laurent~\cite{EhrhardLaurent08}. In these nets, which are graphical objects,
parallel compositions appear as complete graphs, and it is clear that more
general graphs (actually, arbitrary graphs) could be encoded as well in the
very same formalism. A graphical approach to \TCCS{}, in the spirit of
interaction nets, will be presented in a forthcoming paper.

\section*{Acknowledgments}
This work has been partly funded by the French ANR project ANR-07-BLAN-0324
\emph{Curry-Howard for Concurrency} (CHOCO) and by the National Science
Foundation of China project NSFC 61161130530.

\bibliographystyle{alpha} \bibliography{../../newbiblio}

\end{document}

%% file: notation.tex
\theorembodyfont{\normalfont}
\theoremstyle{plain}

\newcommand{\proofitem}[1]{\paragraph{\mdseries\textit{#1}}}

\newcommand{\Beginproof}{\proofitem{Proof.}}
\newcommand{\Endproof}{
  \ifmmode 
  \else \leavevmode\unskip\penalty9999 \hbox{}\nobreak\hfill
  \fi
  \quad\hbox{$\Box$}
  \par\medskip}

\renewcommand{\phi}{\varphi}
\renewcommand\epsilon{\varepsilon}

\newcommand{\Implies}{\Rightarrow}
\newcommand\Equiv{\Leftrightarrow}
\newcommand{\St}{\mid}

\newcommand\cB{\mathcal{B}}

\newcommand\cE{\mathcal{E}}

\newcommand\cI{\mathcal{I}}

\newcommand\cQ{\mathcal{Q}}
\newcommand\cR{\mathcal{R}}
\newcommand\cS{\mathcal{S}}
\newcommand\cT{\mathcal{T}}
\newcommand\cU{\mathcal{U}}
\newcommand\cV{\mathcal{V}}

\newcommand\cX{\mathcal{X}}

\newcommand\Part[1]{{\cal P}({#1})}

\newcommand\Myleft{}
\newcommand\Myright{}

\newcommand\Web[1]{\Myleft|{#1}\Myright|}

\newcommand\IPlus{\oplus}
\newcommand\Plus[2]{{#1}\IPlus{#2}}

\newcommand\Locun[1]{1^J}

\newcommand\Isom\simeq

\newcommand\Comp{\mathrel\circ}

\newcommand\Id{\operatorname{Id}}

\newcommand\Nat{{\mathbf{N}}}

\newcommand\List[3]{#1_{#2},\dots,#1_{#3}}

\newcommand\Subst[3]{{#1}\left[{#2}/{#3}\right]}

\newcommand\Substbis[2]{{#1}\left[{#2}\right]}

\newcommand\Transp[1]{{}^{\mathrm{t}}\!{#1}}

\newcommand\Rel[1]{\mathrel{#1}}

\newcommand\Redst[1]{\mathop{\mathsf{Red}}}

\newcommand\Msetofsubst[1]{\bar F}

\newcommand\TCCS{\textsf{CCTS}}
\newcommand\CCS{\textsf{CCS}}

\newcommand\PREFI[2]{{#1}\cdot{#2}}
\newcommand\PREF[2]{{#1}\cdot{#2}}
\newcommand\PREFO[2]{{\overline{#1}}\cdot{#2}}
\newcommand\PAR[2]{#1\langle#2\rangle}
\newcommand\PLUS{+}
\newcommand\ZERO{0}
\newcommand\FIX[2]{\mu#1\cdot#2}
\newcommand\RESTR[2]{#1\setminus #2}
\newcommand\PARFULL[2]{#1\mid#2}
\newcommand\PAREMPTY[2]{#1\oplus#2}
\newcommand\EMPTY{\epsilon}

\newcommand\Transtau[1]{\overset\tau{\underset{#1}\longrightarrow}}
\newcommand\Transtautr[1]{\overset{\tau*}{\underset{#1}\longrightarrow}}

\newcommand\Translabs[4]{\underset{#3}{\overset{#4:#1\cdot(#2)}\longrightarrow}}

\newcommand\CCStau{\overset\tau{\longrightarrow}}
\newcommand\CCStautr{\overset{\tau*}{\longrightarrow}}
\newcommand\CCSlab[1]{\overset{#1}{\longrightarrow}}

\newcommand\Wtranslabs[6]{\overset{{#2:#1\cdot(#3)}}%
{\underset{#4,#5,#6}{\Longrightarrow}}}

\newcommand\Coh[1]{\frown_{#1}}
\newcommand\Locations{\mathsf{Loc}}
\newcommand\Variables{\cV}

\newcommand\Langofstate{\mathsf L}
\newcommand\Lang[2]{\Langofstate(#1,#2)}

\newcommand\ProcofTA[2]{\langle#1\rangle_{#2}}

\newcommand\ProcofT[1]{\overline{#1}}

\newcommand\Emptyfun\emptyset

\newcommand\Barset[1]{\bar{#1}}

\newcommand\Cansum[1]{\operatorname{\mathsf{cs}}({#1})}

\newcommand\Intred{\mathord\to}
\newcommand\Intredtr{\mathord\to^*}

\newcommand\Cosymb[1]{\bar{#1}}

\newcommand\Processes{\mathsf{Proc}}
\newcommand\Barb[2]{#1\downarrow_{#2}}

\newcommand\Dummy\tau
\newcommand\Wbarbbis{\overset{\bullet}\approx}
\newcommand\Wbarbcong{\cong}
\newcommand\Locof[1]{\Web{#1}}

\newcommand\Wbisim\approx

\newcommand\Gplus[1]{\oplus_{#1}}
\newcommand\Contrel[1]{#1^+}

\newcommand\Casename[1]{(\textbf{#1})}

\newcommand\Coname[1]{\overline{#1}}

\newcommand\Subcasepar{\smallbreak\noindent$\RHD$\ }

\newcommand\CCSbisim[1]{{#1}^{0}}

\newcommand\CCSlocrel[1]{\widehat{#1}}